\documentclass[journal]{IEEEtran}
\IEEEoverridecommandlockouts

\usepackage{amsmath,amssymb,amsfonts}
\usepackage{algorithm}
\usepackage{algorithmicx}
\usepackage{algpseudocode}
\usepackage{setspace}
\usepackage{bm}
\usepackage{graphicx}
\usepackage{textcomp}
\usepackage{xcolor}
\usepackage{tikz}
\usetikzlibrary{spy}
\usepackage{pgfplots}
\pgfplotsset{compat=1.18}
\usetikzlibrary{quantikz2}
\usepackage{float}
\usepackage{subcaption}
\usepackage{mathrsfs}
\usepackage[T1]{fontenc}
\usepackage{verbatim}
\usepackage{hyperref}
\usepackage{makecell}
\usepackage{booktabs}
\usepackage{array} 
\usepackage{comment}
\usepackage{amsthm}
\usepackage{graphbox} 
\usepackage{adjustbox}
\usepackage{balance}
\usepackage{multirow}
\usepackage[capitalize]{cleveref}
\usepackage{footnote}
\usepackage{bm}
\usepackage{threeparttable}
\usepackage{caption}
\usepackage{lipsum}
\usepackage{physics}
\usepackage{braket}
\usepackage{stfloats}
\usepackage{placeins}
\usepackage{cuted}
\usepackage{multicol}
\usepackage[top=0.71in, bottom=1in, left=0.7in, right=0.7in]{geometry}
\def\BibTeX{{\rm B\kern-.05em{\sc i\kern-.025em b}\kern-.08em
    T\kern-.1667em\lower.7ex\hbox{E}\kern-.125emX}}

\newtheorem{prop}{Proposition}

\newtheorem{cor}{Corollary}
\theoremstyle{remark}

\theoremstyle{definition}

\newcommand{\eqdef}{\stackrel{\triangle}{=}}

\newtheorem{remark}{Remark}
    
\begin{document}

\title{Entanglement Generation During Distribution \\
via Spatial Superposition}

\author{Claudio Pellitteri,  Rajiuddin Sk, Marcello~Caleffi,~\IEEEmembership{Senior~Member,~IEEE},\\
and Angela~Sara~Cacciapuoti,~\IEEEmembership{Senior~Member,~IEEE}
    \thanks{The authors are with the \href{www.quantuminternet.it}{www.QuantumInternet.it} research group, University of Naples Federico II, Naples, 80125 Italy.}
    \thanks{A preliminary conference version of this work has been accepted in the Proc. of ICC'26 \cite{SkPelCal-26}. This work has been funded by the European Union under Horizon Europe ERC-CoG grant QNattyNet, n.101169850. 
    Views and opinions expressed are however those of the author(s) only and do not necessarily reflect those of the European Union or the European Research Council Executive Agency. Neither the European Union nor the granting authority can be held responsible for them.}}

\maketitle

\begin{abstract}
The exploitation of quantum coherence at the level of propagation represents a powerful paradigm for quantum communication networks. In this work, we show that the coherent superposition of spatially distinct communication links enables entanglement generation inherently during distribution. Specifically, separable quantum states can be deterministically transformed into entangled states, when the noisy communication links they traverse are coherently superposed. Contrary to the conventional view of noise as a detrimental effect, we demonstrate that quantum noise itself can be transformed into a constructive resource for entanglement generation for both bipartite and multipartite entanglement. Given the practical feasibility of implementing spatial superposition in interferometric setups, our approach provides a feasible method for distributed entanglement engineering, opening new directions for quantum communication and networked quantum technologies.
\end{abstract}


\section{Introduction}
\label{sec:1}
Robust bipartite and multipartite entanglement act as a indispensable resource to design architectures and communication protocols for the future Quantum Internet \cite{CalCac-26,CacCalMet-20,CacCal-26}. However, one of the main challenges in generating and distributing entanglement in large-scale quantum networks is the decoherence that arises from system–environment interactions, leading to rapid degradation of entanglement resources \cite{BucVivCar-08}. The dynamics of entanglement decay and its distribution under various quantum noise models have been extensively investigated since the early development of quantum information theory \cite{HorHorHor-01,BelCom-08,MaSunWan-12}. Quantum repeaters have been proposed to extend the range of entanglement distribution \cite{AzEcSo-23}. Complementary approaches have also been developed for the generation and preservation of quantum entanglement, including entanglement purification \cite{PanSimBru-01,YanZhoZho-20}, weak measurements \cite{KimLeeKim-12}, local filtering \cite{HuaXinYan-14}, and various error-mitigation techniques \cite{CaiBabBen-23}. 

While these approaches aim at counteracting the detrimental effects of quantum noise, they fundamentally treat noise as an adversarial phenomenon to be suppressed or corrected.

\vspace{1mm}
This perspective prompts a fundamental question: \textit{can quantum noise itself, inevitably affecting a quantum state as it propagates through a quantum channel, be harnessed as a constructive resource for entanglement generation?} 

In this work, we answer this counterintuitive question, by showing that the appropriate exploitation of quantum coherence at the level of propagation enables the deterministic generation of bipartite and multipartite entangled states from initially separable input-states. The central idea is that a quantum particle is not constrained to follow a single, well-defined trajectory. Rather, it can propagate along multiple trajectories simultaneously in coherent superposition. As a consequence, not only the particle’s internal degrees of freedom, but also the very instantiation of the communication channels through which it travels, can acquire a genuinely quantum character. In this framework, the instantiation of communication channels\footnote{In the following, the terms communication link and channel are often used interchangeably, since each communication link is associated with an underlying quantum channel describing the physical evolution of a quantum particle traversing that link.} is no longer a fixed classical structure, but it can instead be coherently superposed. 

\begin{figure*}
    \centering
    \includegraphics[width=0.99\linewidth]{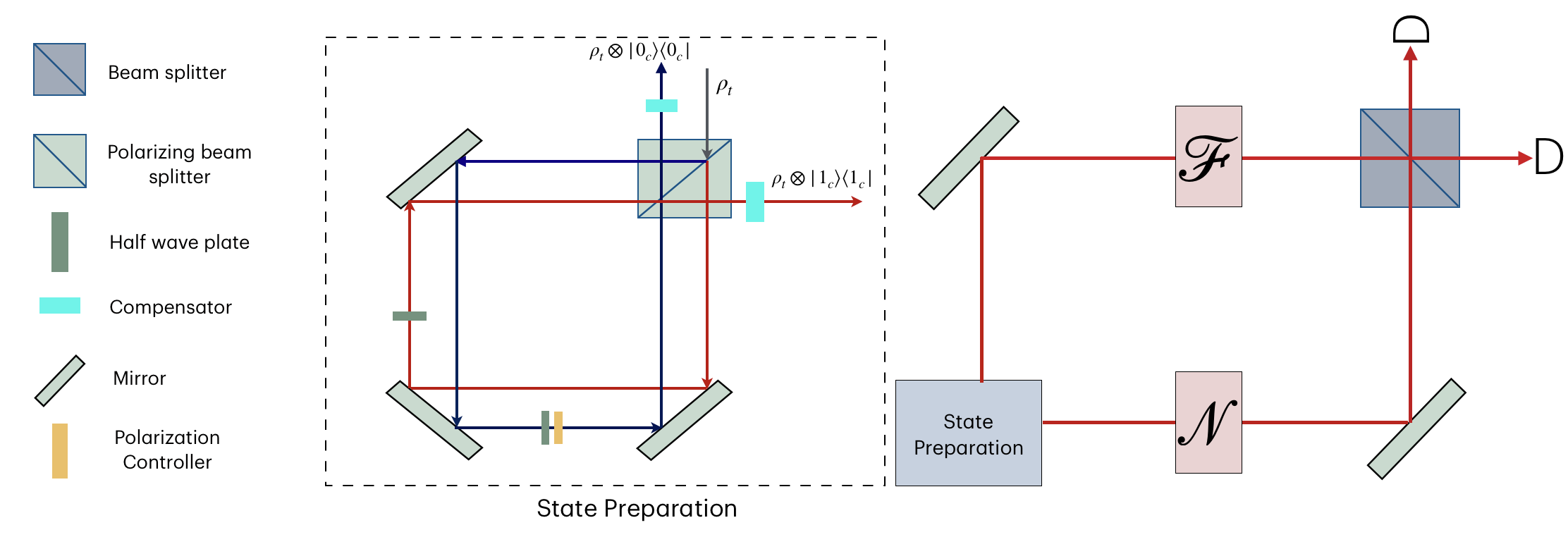}
    \caption{Photonic setup for realizing the spatial superposition of two communication links, described by the quantum channel maps $\mathcal{F}(\cdot)$ and $\mathcal{N}(\cdot)$. 
    The setup consists of two interferometers: a Sagnac interferometer for state preparation and a Mach-Zehnder interferometer for generating the superposition of the two links. The photon source enters the Sagnac interferometer, where two HWPs are rotated in a correlated manner; for a control qubit in the state $\ket{+}$, both HWPs are oriented at $90^{\circ}$ with respect to the horizontal axis. Two compensators (each consisting of a HWP and a QWP) are placed at the two output ports of the Sagnac interferometer to correct path-dependent polarization effects. The two output beams are then injected into the two arms of the Mach-Zehnder interferometer and interfere at the final beam splitter \cite{PaLuNo-23}.}
    \label{fig:01}
    \rule{\linewidth}{\arrayrulewidth}
\end{figure*}

As shown in \cite{ChiKri-19}, such superpositions of trajectories can arise in two conceptually distinct ways. The first corresponds to a superposition of spatial trajectories, where the particle coherently propagates along distinct communication links. The second involves a superposition of the orders in which the particle traverses two or more links, a scenario commonly referred to in the literature as ``quantum switch''. The experimental realization of a quantum switch is very challenging \cite{WeiTisZha-19,RubRozWal-17}. And, there is an ongoing debate concerning the physical interpretation of existing demonstrations, in particular whether they genuinely implement an indefinite causal order or should rather be regarded as simulations of the quantum switch \cite{HamKabBru-25}. On the other hand, superposition of spatial links can be implemented with significantly lower experimental overhead, for instance by means of standard interferometric setups or through architectures based on discrete quantum walks. In addition, the underlying physical picture of spatial superposition is well understood and widely accepted \cite{AbbWecBra-20}. A comparative overview of these two frameworks is presented in Table \ref{tab:01}. 
In \cref{fig:01}, a photonic setup for the the superposition of two communication links is displayed. In this setup, two degrees of freedom of a quantum particle, polarization and path, are employed: the polarization serves as the input state, while the path mode encodes the control qubit.  

Motivated by this near-term experimental feasibility, we focus on the use of spatial superposition for entanglement generation. Specifically, we exploit the coherent superposition of communication links to generate both bipartite and multipartite entanglement. Our findings reveal that by coherently superposing the actions of noisy channels, separable input-states can be deterministically transformed into maximally entangled states. These findings open new perspectives on noise-assisted entanglement engineering and quantum network design.

\begin{table*}[t]
\caption{Comparison between quantum switch and spatial superposition frameworks.}
\renewcommand{\arraystretch}{1.2} 
\setlength{\tabcolsep}{8pt}       
\begin{tabular}{p{4cm}p{6cm}p{6cm}}
\hline\hline
\textit{Feature} & \textit{Quantum Switch} & \textit{Spatial Superposition } \\
\hline
Physical mechanism 
& Superposition of alternative causal orders among the communication links 
& Coherent superposition of distinct (in space) communication links \\
\hline
Main operational principle 
& Indefinite causal order between quantum operations modeling quantum channel maps, underlying the communication links
& Interference between alternative quantum operations modeling quantum channel maps, underlying the communication links \\
\hline 
Experimental feasibility 
& Challenging 
& Comparatively simple \\
\hline\hline
\end{tabular}
\label{tab:01}
\end{table*}
\subsection{Related Works}
The quantization at the propagation level \cite{OreCosBru-12,RozStrWal-24,GisLinPop-05,GueRubBru-19,PelCalCac-25}, as realized through quantum switch \cite{EblSalChi-18} and coherent control of spatially different links \cite{AbbWecBra-20}, has been shown to provide significant advantages in quantum communication \cite{EblSalChi-18,CalCac-20,RubRozChi-21,ChaCalCac-21,SimCalIll-26,CalSimCac-23} and quantum computation \cite{ChiGiaVal-13,AraMatPhi-17,ApaBisPer-24}.  
Only a limited number of works have investigated entanglement generation within the framework of indefinite causal order. In \cite{KoCaCa-23}, multipartite entanglement has been generated via an indefinite causal order of unitary gates. However, this approach is restricted to purely unitary operations and requires additional constraints on the allowed unitaries. In \cite{LiWeWe-25}, a scheme is proposed to generate multipartite entanglement between distant parties by combining the indefinite causal order framework with preshared entanglement and local unitary operations.

Very recently, \cite{ChaShiDur-2025} proposed a quantum networking framework based on spatial superposition to dynamically establish high-fidelity entanglement between different nodes of a network. The objective of that work is to mitigate the effect of noise on the entanglement distribution, rather than to generate entanglement. Furthermore, the proposed protocol requires a comparatively large amount of physical and operational resources than ours. 

Another related framework is the discrete-time quantum walk, which constitutes the quantum analogue of classical random walk and has been widely investigated as a versatile tool for quantum information processing. Its use for entanglement generation has been explored in various scenarios, including entanglement between two walkers and between different degrees of freedom of a single walker. In \cite{viAmRi-13,OmPaBo-06,LiSha-21}, bipartite and multipartite entanglement has been systematically demonstrated in qubit and qudit systems. However, it is important to note that the discrete-time quantum walk  can be interpreted as a specific realization of the more general framework based on the coherent superposition of spatially different links. From this perspective, the dynamics of quantum walks can be emulated within the spatial superposition framework, which therefore constitutes a more general model. In Table~\ref{tab:02}, we present a brief comparative overview of these two approaches.

Despite this conceptual connection, entanglement generation based on the explicit exploitation of spatial superposition has not yet been investigated. To the best of our knowledge, this work constitutes the first demonstration of entanglement generation through this mechanism in a fully noisy setting. Unlike existing approaches, which typically aim to preserve pre-existing entanglement against noise, our framework instead harnesses noisy dynamics as a resource, enabling entanglement generation intrinsically during propagation.
\begin{table*}[]
\caption{Comparison between spatial superposition and quantum walk frameworks.}
    \centering
    \renewcommand{\arraystretch}{1.2} 
\setlength{\tabcolsep}{8pt}      
\begin{tabular}{p{4cm}p{6cm}p{6cm}}
\hline\hline
\textit{Feature} & Spatial Superposition  & Quantum Walk \\
       \hline\hline
       Classical counterpart & None & Classical random Walk \\
       \hline
       Physical mechanism & Coherent superposition of spatially distinct communication links  & Coherent evolution of a quantum walker state conditioned on a coin state\\
       \hline
        Operational principle & Interference between alternative links & Stepwise unitary evolution driven by coin and shift operators \\
       \hline
       Supported dynamics for entanglement generation & Applicable to both noisy and noiseless scenarios & Typically limited to noiseless (unitary) dynamics\\
       \hline
       Implementation & Interferometric setup & Lattice-based architectures\\
       \hline
       Generality & Supports coherent superposition of arbitrary quantum operations (unitary and non-unitary) in a single step & Corresponds to specific unitary dynamics. It can be realized within the spatial superposition framework 
       \\
         \hline\hline
    \end{tabular}
    \label{tab:02}
\end{table*}

The remainder of the paper is organized as follows. Sec.~\ref{sec:2} reviews the theoretical formalism of spatial superposition and explicitly establishes its connection with discrete-time quantum walks. Sec.~\ref{sec:3} presents the generation of Bell, GHZ and W states from separable input-states via superposition of unitary channels. Sec.~\ref{sec:4} analyzes the framework under noisy and Sec.~\ref{sec:5} concludes the paper.

\section{Spatial Superposition of Links}
\label{sec:2} 

\begin{figure}
    \centering
    \includegraphics[width=0.7\linewidth]{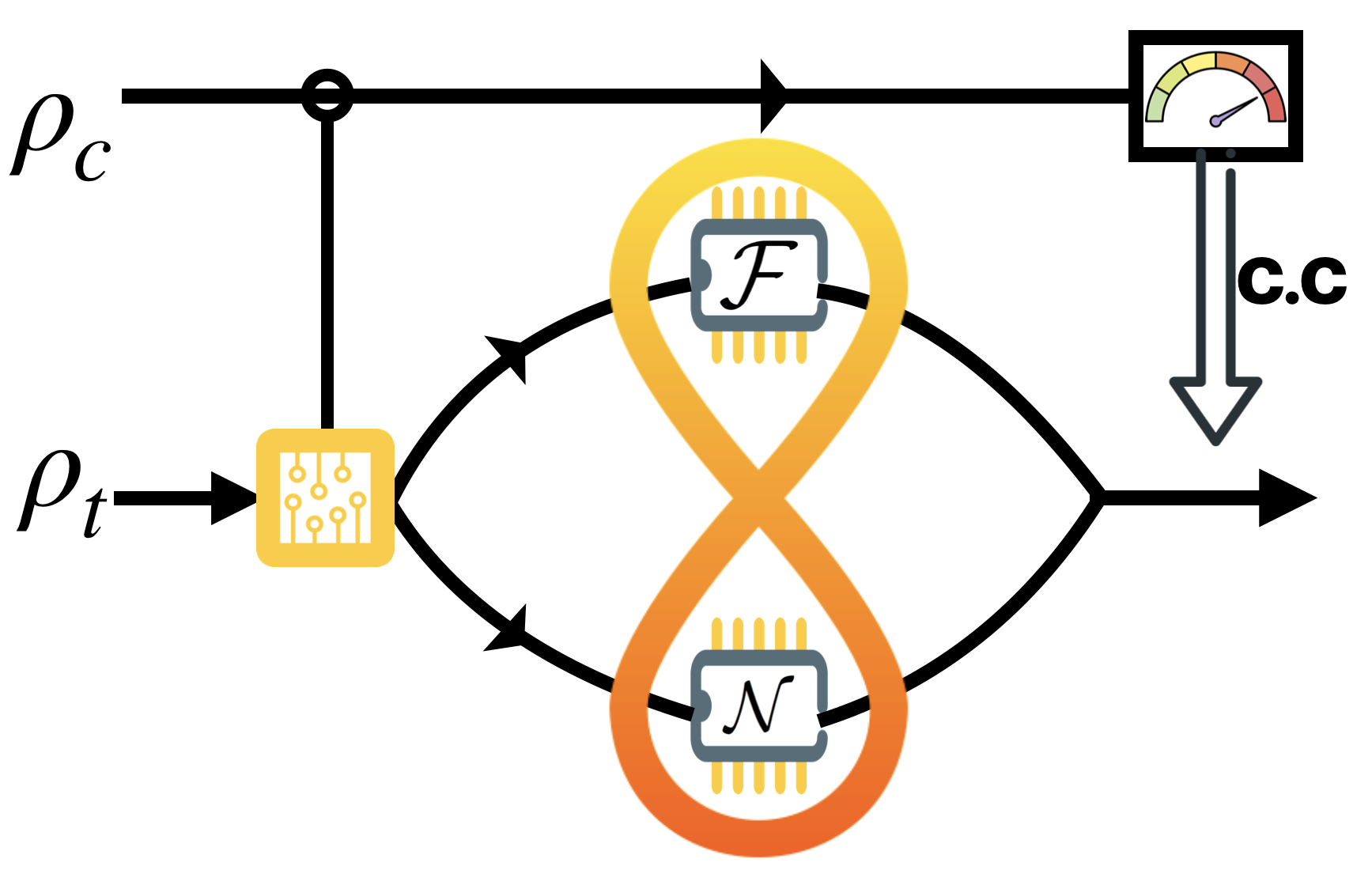}
    \caption{Schematic diagram illustrating the superposition of two distinct quantum communication links, associated with the quantum channels $\mathcal{F}(\cdot)$ and $\mathcal{N}(\cdot)$.}
    \label{fig:02}
    \rule{\linewidth}{\arrayrulewidth}
\end{figure}

The mathematical formalism underlying the superposition of spatially distinct communication links has been developed in \cite{ChiKri-19}. This formalism is based on extending the Hilbert space associated to the considered quantum system by introducing the so-called vacuum state, which accounts for the case a given channel is not traversed by any particle. This extension combined with the inclusion of a quantum-control system -- realized as a qubit or, more generally, as a qudit (see \cref{fig:02} and \cref{fig:03}) -- enables the description of a particle propagating either through a single channel or through their coherent superposition. Mathematically, the superposition of two unitary quantum operations, $U_1$ and $U_2$, acting on a target state $\rho_t$ and controlled by a quantum system in the state $\rho_c$, is described as follows:
\begin{equation}
\label{eq:01}
\mathcal{S}(U_1,U_2)(\rho_t)=S(\rho_t\otimes\rho_c)S^\dagger,
\end{equation}
with $S=U_1\otimes \ket{0_c}\bra{0_c}+U_2 \otimes\ket{1_c}\bra{1_c}$. When the system instead evolves under a coherent superposition of quantum channels, the corresponding map is as follows:
\begin{equation}
\label{eq:02}
\mathcal{S}(\mathcal{F},\mathcal{N})(\rho_t)=\sum_{ij}S_{ij}(\rho_t\otimes\rho_c)S^\dagger_{ij}
\end{equation}
with $S_{ij}$ given by:
\begin{equation}
\label{eq:03}
   S_{ij}=\beta_j F_i \otimes \ket{0_c}\bra{0_c} +\alpha_i N_j\otimes \ket{1_c}\bra{1_c}.
\end{equation}
Here, the channel maps $\mathcal{F}(\cdot)$ and $\mathcal{N}(\cdot)$ are characterized by the Kraus operators $\{F_i\}$ and $\{N_j\}$, and by the corresponding vacuum amplitudes $\{\alpha_i\}$ and $\{\beta_j\}$. These amplitudes satisfy the normalization conditions $\sum_i \alpha_i^2=1$ and $\sum_j \beta_j^2=1$ \cite{ChiKri-19}. Eq.~\cref{eq:01} can be regarded as a special case of \cref{eq:02}, in the absence of environmental interactions.

\subsection{Discrete-Time Quantum Walk as a special case}
The discrete-time quantum random walk (DTQW) is the quantum counterpart of the classical random walk \cite{VenAnd-2012}. In the classical setting, the simplest random walk describes the motion of a particle on a one-dimensional lattice, where at each step the particle moves either left or right according to the outcome of a binary random variable, typically modeled as a coin toss. Translating this dynamics into the quantum framework requires introducing two distinct Hilbert spaces: an $N$-dimensional position space $H_p$, spanned by basis states $\{\ket{i}\}$ with $i=1,\ldots,N$, describing the walker's position on a one-dimensional lattice, and a two-dimensional coin space, spanned by $\{\ket{0},\ket{1}\}$, encoding the directional degree of freedom. The evolution of the quantum walker is governed by the unitary operator
\begin{equation}
    \label{eq:04}
    U = T (I \otimes C),
\end{equation}
where $C$ denotes a unitary operator acting on the coin state, $I$ is the identity operator on the position space, and $T$ is the conditional shift operator defined as $T = \sum_i \ket{i+1}\bra{i} \otimes \ket{0}\bra{0} + \sum_i \ket{i-1}\bra{i} \otimes \ket{1}\bra{1}$.
A striking departure from classical behavior lies in the non-zero average displacement of the quantum particle \cite{Kempe2003}. Furthermore, when a balanced coin operator is employed and the initial coin state is prepared in a coherent superposition of its two basis states, the resulting probability distribution remains symmetric. This symmetry arises from the interference of multiple paths simultaneously explored by the walker along the lattice.
The DTQW can be alternatively interpreted as a specific realization of a more general framework based on the the coherent superposition of quantum trajectories.
In this framework, the dynamics is described in terms of two subsystems: a control system and a target system. A natural correspondence can be established, by identifying the coin degree of freedom with the control system and the position degree of freedom with the target system. 
Under this mapping, a direct comparison between the operators $S$ in \eqref{eq:01} and $U$ in \eqref{eq:04} reveals that, for specific choices of the unitaries $U_1$ and $U_2$, the dynamics induced by $S$ on a given input state coincides with that of a standard DTQW defined on a two-vertex graph.

\section{Entanglement  Generation via Spatial Superposition: Ideal Regime}
\label{sec:3}
Here, we investigate entanglement generation in the ideal regime, i.e., in the absence of quantum noise.  We first analyze the bipartite case, by showing that Bell states can be generated from separable two-qubit input states through the spatial superposition framework, as formalized in Prop.~\ref{prop:01}. Then we extend the framework to the multipartite setting, by considering the generation of GHZ and W states, as formalized in Props.~\ref{prop:02} and \ref{prop:03}, respectively.
\begin{prop}
\label{prop:01}
A two-qubit separable state $\rho_t=\ket{00}\bra{00}$ evolving under a coherent superposition of bit-flip $X\otimes X$ and phase-flip $Z \otimes Z$ operations yields a maximally entangled two-qubit state (Bell pair), for any measurement outcome of the control qubit.
\begin{proof}
   Please refer to Appendix~\ref{Appendix_A}.
\end{proof}
\end{prop}
As shown in Appendix~\ref{Appendix_A}, measuring the control qubit in $\ket{+_c}$ yields the Bell-state output $\ket{\Phi^+} = \frac{1}{\sqrt{2}}(\ket{00} + \ket{11})$, while the $\ket{-_c}$ outcome yields $\ket{\Phi^-} = \frac{1}{\sqrt{2}}(\ket{00} - \ket{11})$, which is locally-unitary equivalent to $\ket{\Phi^+}$. Interestingly, for an arbitrary separable input-state of the form $\rho_t= \ketbra{\psi_t}{\psi_t}$, with $\ket{\psi_t}=(a\ket{0}+b\ket{1})^{\otimes 2} $, the resulting output states are still maximally entangled Bell states, regardless of the relative amplitude of the input state.

Beyond bipartite entanglement, multipartite states, where entanglement is shared among more than two parties, also play a key role in quantum networking. Among them, two prominent classes that enable different network functionalities \cite{IllCalMan-20} are the GHZ and W states \cite{CheChe-06, MazCalCac-25, DurVidCir-00}, both exhibiting genuine multipartite entanglement, i.e., being non-factorizable across any partition of the Hilbert space.
\subsection{Generation of GHZ states}
\label{sec:3.1}
The generation of GHZ states from separable inputs via spatial superposition naturally extends the Bell-state framework. Indeed, $\ket{\Phi^+}$ can be regarded as the 2-qubit instance of a GHZ state. 
\begin{prop}
\label{prop:02}
    An n-qubit separable state $\rho_t=\ket{0}^{\otimes n}\bra{0}^{\otimes n}$ evolving under a coherent superposition of bit-flip $X^{\otimes n}$ and phase-flip $Z^{\otimes n}$ operator deterministically yields a $n$-qubit GHZ state, up to a relative phase determined by the control measurement outcome.
\begin{proof}
    Please refer to Appendix~\ref{Appendix_A}.
\end{proof}
\end{prop}
Accordingly, a $n$-qubit GHZ state can be deterministically generated by coherently superposing $X^{\otimes n}$ and $Z^{\otimes n}$.
\subsection{Generation of W states}
\label{sec:3.2}
\begin{figure}
    \centering
    \includegraphics[width=0.75\linewidth]{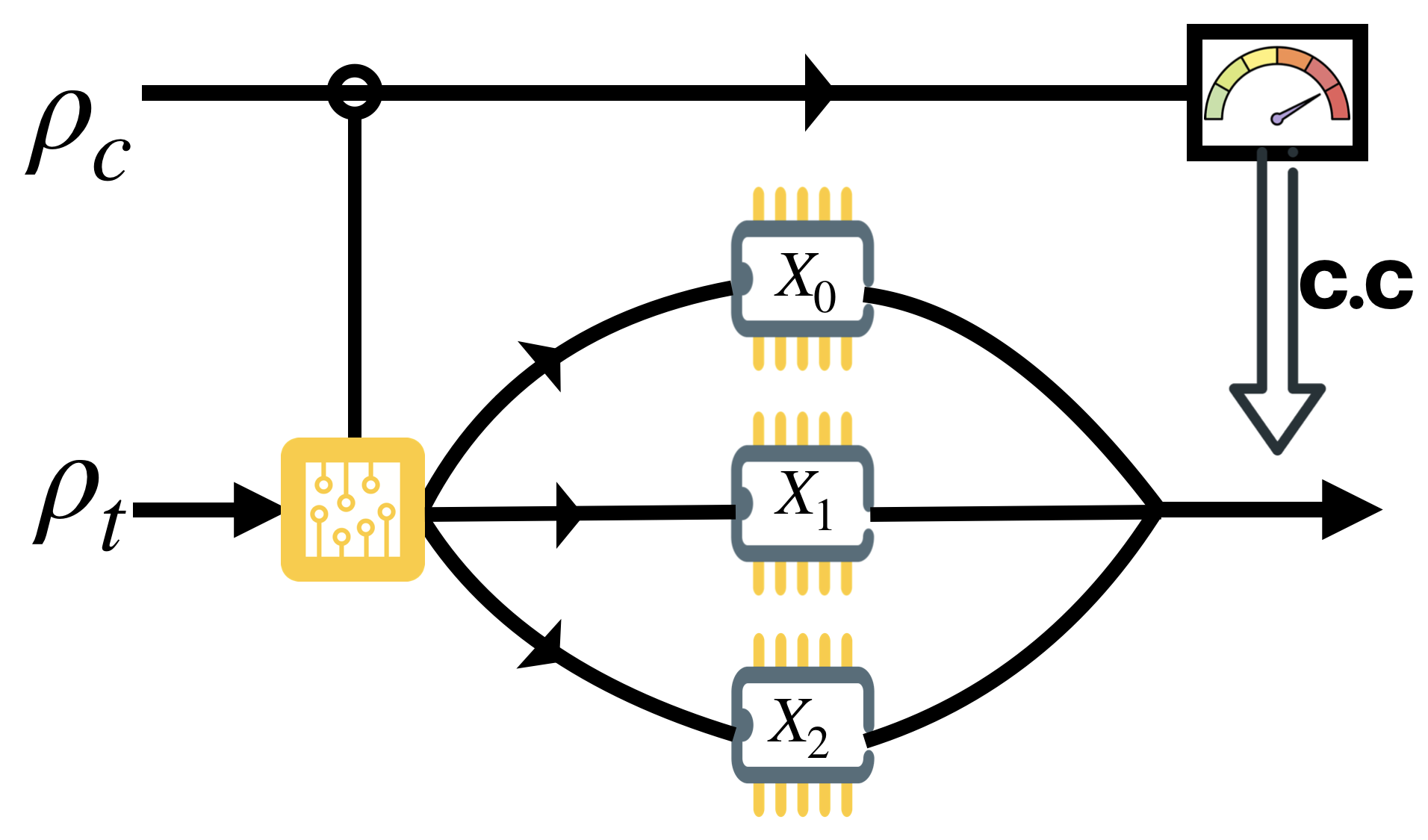}
    \caption{Schematic diagram illustrating the superposition of three links used to generate a $3$-qubit W state. The control qubit is prepared in the state $\ket{e_+}=\frac{1}{\sqrt{3}}(\ket{0}+\ket{1}+\ket{2})$, while the target in $\rho_t=\ket{000}\bra{000}$. The applied operations are $X_0\eqdef X\otimes I \otimes I$, $X_1\eqdef I\otimes X \otimes I$ and $X_2\eqdef I\otimes I \otimes X$.}
    \label{fig:03}
    \rule{\linewidth}{\arrayrulewidth}
\end{figure}

The generation of W states is considerably less straightforward and requires the coherent superposition of more than two operations, as illustrated in \cref{fig:03} and formalized in the following proposition.
\begin{prop}
\label{prop:03}
    An $n$-qubit separable state $\rho_t=\ket{0}^{\otimes n}\bra{0}^{\otimes n}$ evolving under a coherent superposition of the $n$ operators $\{X_i\}$ deterministically yields an $n$-qubit W state, up to a local unitary depending on the outcome of the control measurement, where
    $X_i \eqdef \mathbb{I}_0\otimes\mathbb{I}_1...\otimes\mathbb{I}_{i-1}\otimes X\otimes\mathbb{I}_{i+1}\otimes...\otimes\mathbb{I}_{n-1}$ acts non-trivially only on the $i$-th qubit.
\begin{proof}
    Please refer to Appendix~\ref{Appendix_B}.
\end{proof}
\end{prop}
Consequently, the generation of an $n$-qubit W state requires the coherent superposition of $n$ unitaries. And, the control system is no longer a qubit, but an $n$-dimensional system (qudit), with computational basis $B=\{\ket{i},i=0,...,n-1\}$. The control is initialized in the state $\ket{\psi}=1/\sqrt{n}\sum^{n-1}_{i=0}\ket{i}$, leading to a coherent superposition of all the unitaries. 

\section{Entanglement  Generation via Spatial Superposition: Noisy Regime}
\label{sec:4}
In this section, we investigate entanglement generation in the presence of quantum noise, first in the bipartite setting and then in the multipartite case.
\subsection{Bipartite Entanglement Generation under Noise}
To capture the most adverse conditions, we consider the superposition of two depolarizing channels $\mathcal{F}(\cdot)$ and $\mathcal{N}(\cdot)$, representing a worst-case noise model due to their fully randomizing effect:
\begin{align}
\label{eq:05}
    &\mathcal{F}(\rho)=(1-p) \rho+\frac{p}{3} (X\otimes X)\rho (X\otimes X)\nonumber\\
    &+\frac{p}{3} (Y\otimes Y)\rho (Y\otimes Y)+\frac{p}{3} (Z\otimes Z)\rho (Z\otimes Z),
    \end{align}
\begin{align}
 \label{eq:06}
   &\mathcal{N}(\rho)=(1-q) \rho+\frac{q}{3} (X\otimes X)\rho (X\otimes X)\nonumber\\
   &+\frac{q}{3} (Y\otimes Y)\rho (Y\otimes Y)+\frac{q}{3} (Z\otimes Z)\rho (Z\otimes Z).
\end{align}
%
As shown in Appendix~\ref{Appendix_C}, the fidelity $\mathrm{fid}$ between the Bell state $\ket{\Phi^+}$ and the $2$-qubit output state, obtained via the spatial superposition in \eqref{eq:02} of the  depolarizing channels in \eqref{eq:05} and \eqref{eq:06}, admits the following closed-form expression:
\begin{equation}\label{eq:07}
    \mathrm{fid}=\sqrt{\frac{C}{D}},
\end{equation}
with
\begin{align}
\label{eq:08}
C &= 3+3\sqrt{(1-p)(1-q)}\,\alpha_{0}\beta_{0} \\\nonumber
& +\sqrt{3p(1-q)}\beta_{0}(\alpha_{1}-\alpha_{2}+\alpha_{3})
+\sqrt{3q(1-p)}\alpha_{0}\\\nonumber&(\beta_{1}-\beta_{2}+\beta_{3})
+\sqrt{pq}(\alpha_{1}-\alpha_{2}+\alpha_{3})(\beta_{1}-\beta_{2}+\beta_{3}), 
\end{align}
and
\begin{align}
\label{eq:09}
D &= 2\Big(3+3\sqrt{(1-p)(1-q)}\,\alpha_{0}\beta_{0}
+\sqrt{pq}\alpha_{3}\beta_{3} \\\nonumber
& +\sqrt{3p(1-q)}\beta_{0}\alpha_{3}+\sqrt{3q(1-p)}\alpha_{0}\beta_{3}\\\nonumber
&+\sqrt{pq}(\alpha_{1}-\alpha_{2})(\beta_{1}-\beta_{2})\Big).
\end{align}
The above expressions show that the fidelity depends both on the noise parameter $p$ and $q$ and on the vacuum amplitudes $\{\alpha_i\}$ and $\{\beta_j\}$. The noise strengths are fixed by the intrinsic properties of the channels, whereas the vacuum amplitudes parametrize the corresponding vacuum extensions. In particular, they encode the relative weights and phases associated with different coherent evolutions in an extended Stinespring representation of the channel, i.e., of the same communication resource\footnote{Physically, the vacuum extension provides a complete description of the communication resource available to the sender and receiver, and it can be experimentally reconstructed via an input–output tomography \cite{ChiKri-19,PaLuNo-23}.}. As discussed in~\cite{ChiKri-19}, the vacuum extensions of a given quantum channel are generally non-unique and form a convex set. Hence, different choices of the vacuum amplitudes correspond to different points in the convex set of admissible vacuum extensions of the same reduced CPTP map. Therefore, while these extensions are mathematically distinct, they all induce the same map at the level of the reduced system. However, they become operationally distinguishable and inequivalent under coherent superposition, since they induce different interference effects.

From an operational perspective, this freedom manifest as the ability to engineer the interference patterns arising from the superposition of different communication paths. This can be accomplished by controlling experimentally accessible parameters that determine the relative amplitudes and phases of the coherent evolutions. In other words, although the vacuum extension is not unique from a mathematical perspective, it is uniquely determined by the physical realization of the communication framework \cite{ChiKri-19}. In this sense, the vacuum amplitudes can be regarded as parameters labeling the admissible extensions of the same channel, among which a given experimental implementation selects a specific one. For instance, in interferometric implementations, such control is naturally achieved by tuning the configuration of the setup, e.g., through beam splitters transmissivities, relative phase shifts, and other path-dependent transformations ruling the amplitudes and phases on the different branches~\cite{PaLuNo-23}. In particular \cite{PaLuNo-23} shows that different interference regimes (phase-coherent and phase-incoherent) can be realized by modifying the interferometric configuration, for instance by introducing a phase modulator in one arm.

This observation is central to our approach, as it enables the optimization of entanglement generation over the space of admissible vacuum extensions. 
Specifically, for given values of the noise parameters $p$ and $q$, the fidelity can be optimized with respect to the vacuum amplitudes. In practice, the noise strengths can be estimated via quantum process tomography, and the corresponding optimal vacuum amplitudes can then be determined. However, in the general case when $p \neq q$, such an optimization does not admit a closed-form solution and must be carried out numerically. To obtain analytical insights, we therefore focus on the symmetric scenario $p=q$. In particular, we consider the most adverse regimes, namely $p = q = 1/2$, corresponding to a zero-capacity channel, and $p = q = 1$, corresponding to a fully depolarizing and entanglement-breaking channel. In these cases, a closed-form characterization can be derived, as formalized in Prop.~\ref{prop:04}, which highlights the key role played by the choice of vacuum extension in determining the entanglement generation capability of the proposed framework.


\begin{figure*}[t]
    \centering
    \begin{subfigure}[t]{0.48\textwidth}
        \centering
        \vspace{0pt}
        \begin{tikzpicture}[spy using outlines={rectangle, magnification=2, size=1.2cm, connect spies}]

\definecolor{darkgray176}{RGB}{176,176,176}
\definecolor{dimgray}{RGB}{105,105,105}
\definecolor{gray}{RGB}{128,128,128}
\definecolor{lightgray204}{RGB}{204,204,204}
\definecolor{navy}{RGB}{0,0,128}
\definecolor{blue}{RGB}{0,0,255}
\definecolor{green}{RGB}{0,180,0}

\begin{axis}[
    width=8cm, height=6cm,
    legend cell align={left},
    legend style={
      at={(0.5,-0.30)},
      anchor=north,
      legend columns=1,
      draw=black,
      fill=white
    },
    tick align=outside,
    tick pos=left,
    grid=both,
    x grid style={lightgray204},
    xlabel={\Large{Probability $p$}},
    xmin=0.0, xmax=1.0,
    xtick style={color=black},
    xtick={0, 0.2, 0.5, 0.8, 1},
    ylabel={\Large{$\mathrm{fid}$}},
    ymin=0.7, ymax=1.0,
    ytick style={color=black},
    ytick={0.7, 0.8, 0.9, 1.0},
    line width=0.65pt
]

\addplot [thick, mark=*, red]
table {
0 0.707107
0.0024536 0.707493 
0.0409084 0.713654 
0.121109 0.727224 
0.204257 0.742429 
0.284231 0.758292 
0.367151 0.776206
0.448938 0.795559 
0.530611 0.816824 
0.61217 0.840326 
0.693616 0.866469 
0.773927 0.89538 
0.857186 0.929339 
0.93727 0.966726 
0.978721 0.988272 
0.989017 0.993887 
0.991763 0.995403
0.995881 0.997693 
0.998627 0.999229 
1 1
};

\addlegendentry{\shortstack{
$\alpha_0=\beta_0=0$\\
$\alpha_1=\beta_2=\beta_3=-\alpha_2=-\alpha_3=-\beta_1=\frac{1}{\sqrt{3}}$
}}
\addplot [thick,mark=square, green]
    table {%
0. 0.707107 
0.05 0.707107
0.1 0.707107 
0.15 0.707107
0.2 0.707107
0.25 0.707107 
0.3 0.707107
0.35 0.707107
0.4 0.707107 
0.45 0.707107 
0.5 0.707107 
0.55 0.707107
0.6 0.707107 
0.65 0.707107
0.7 0.707107 
0.75 0.707107 
0.8 0.707107 
0.85 0.707107 
0.9 0.707107 
0.95 0.707107
1. 0.707107
};
\addlegendentry{$\alpha_0=\alpha_1=\alpha_2=\alpha_3=\beta_0=\beta_1=\beta_2=\beta_3=\frac{1}{\sqrt{4}}$}

\addplot [thick, mark=diamond, blue]
table {
0 0.707174 
0.0027603 0.732689 
0.00981434 0.756531 
0.0202937 0.779526 
0.0421502 0.813625 
0.142481 0.904871 
0.305546 0.975962 
0.356568 0.9874 
0.386907 0.992324 
0.39828 0.993835 
0.410941 0.995311 
0.429331 0.99708 
0.439747 0.99789 
0.451595 0.998648 
0.460231 0.999092 
0.472677 0.999574 
0.48074 0.999789 
0.486322 0.999894 
0.491564 0.99996 
0.494908 0.999985 
0.497645 0.999997 
0.500077 1 
0.502509 0.999996 
0.505246 0.999985 
0.509502 0.999949 
0.514779 0.999878 
0.521376 0.999746 
0.530919 0.99947 
0.541692 0.999041 
0.555647 0.998302 
0.571987 0.997178 
0.593231 0.995306 
0.672563 0.984299 
0.83619 0.940492 
0.920263 0.901936 
0.942542 0.887976 
0.958665 0.87597 
0.969426 0.866544 
0.983526 0.851106 
0.99039 0.841037 
0.995881 0.82981 
0.99897 0.819193 
1 0.808655
};

\addlegendentry{\shortstack{
$\alpha_0=-\beta_0=-\frac{1}{\sqrt{2}}$\\
$\alpha_1=\beta_2=\beta_3=-\alpha_2=-\alpha_3=-\beta_1=\frac{1}{\sqrt{6}}$
}}

\end{axis}
\end{tikzpicture}
        \caption{Depolarizing Channels.}
        \label{fig:04_a}
    \end{subfigure}
     \hfill
    \begin{subfigure}[t]{0.48\textwidth}
        \centering
        \vspace{0pt}
        \begin{tikzpicture}[spy using outlines={rectangle, magnification=2, size=1.2cm, connect spies}]

\definecolor{darkgray176}{RGB}{176,176,176}
\definecolor{dimgray}{RGB}{105,105,105}
\definecolor{gray}{RGB}{128,128,128}
\definecolor{lightgray204}{RGB}{204,204,204}
\definecolor{navy}{RGB}{0,0,128}
\definecolor{blue}{RGB}{0,0,255}
\definecolor{green}{RGB}{0,180,0}

\begin{axis}[
    width=8cm, height=6cm,
    legend cell align={left},
    legend style={
      fill opacity=0.8,
      draw opacity=1,
      text opacity=1,
      at={(0.5,-0.28)},     
        anchor=north,         
        legend columns=1,     
        draw=none,            
        legend cell align=left,
        align=left,
        draw=black,          
        fill=white, 
      /tikz/every even column/.append style={column sep=0.1cm}
    },
    tick align=outside,
    tick pos=left,
    grid=both,
    x grid style={lightgray204},
    xlabel={\Large{Probability $p$}},
    xmin=0.0, xmax=1.0,
    xtick style={color=black},
    xtick={0, 0.2,0.5,0.8,1},
    xticklabels={
      \(\displaystyle {\ensuremath{0}}\),
      \(\displaystyle {\ensuremath{0.2}}\),
      \(\displaystyle {0.5}\),
      \(\displaystyle {\ensuremath{0.8}}\),
      \(\displaystyle {1}\)
    },
    ymode=log,
    log basis y={10},
    y grid style={darkgray176},
    ylabel={\Large{$\mathrm{fid}$}},
    ymin=0.7, ymax=1.0,
    ytick style={color=black},
    ytick={0.5,0.7 ,0.8,0.9,1.0},
    line width=0.65pt,
    yticklabels={
      \(\displaystyle {0.5}\),
      \(\displaystyle {0.7}\),
      \(\displaystyle {0.8}\),
      \(\displaystyle {0.9}\),
      \(\displaystyle {1.0}\)
    }
   ]
    
    \addplot [thick,mark=*, red]
    table {%
0 0.7071067884019228
0.0196286612015304 0.7140128364397696 
0.16346277092346467 0.7627131737827348
0.305545518858404 0.8079435372779473
0.44893812949614426 0.8511574852799403
0.5926201946734201 0.8923620886930987
0.7345508980637009 0.9312762474324416 
0.8767710559935172 0.9687030133104565
1. 1.
};
    \addlegendentry{$\alpha_0=\beta_0=0$ and $\alpha_1=\beta_1=1$}
    \addplot [thick, mark=square, green]
    table {%
0 0.707124
0.00368039 0.714497
0.0122679 0.720935
0.0412188 0.733489
0.142481 0.759414
0.345986 0.794525
0.55031 0.823262 
0.754434 0.849345 
0.918441 0.867802
0.93727 0.869418
0.950449 0.870368
0.963276 0.871065 
0.968811 0.871263 
0.97619 0.871383 
0.982839 0.871279
0.987644 0.871002 
0.999657 0.867275 
1. 0.866036
};
    \addlegendentry{$\alpha_0=\beta_0=\alpha_1=\beta_1=\frac{1}{\sqrt{2}}$}
    \addplot [thick,mark=diamond,blue]
    table {%
0 0.707157
0.003067 0.728553
0.0104277 0.749415
0.0249486 0.777237
0.0607779 0.826034
0.14 0.898359
0.243493 0.955532 
0.348631 0.986456
0.377029 0.991358
0.39828 0.99423
0.413568 0.995906
0.429944 0.997359
0.444036 0.998
0.49278 0.999974 
0.5 1
0.508286 0.999966 
0.562319 0.99817
0.58214 0.996876
0.633371 0.992102
0.814805 0.960103 
0.938588 0.920312 
0.958357 0.910647
0.969426 0.904207 
0.983869 0.893692 
0.992106 0.885342
0.997254 0.877397
};
\addlegendentry{$\alpha_0=-\frac{1}{\sqrt{2}}$ and $\beta_0=\alpha_1=\beta_1=\frac{1}{\sqrt{2}}$}

 \addplot [thick, red, dotted]
 table {%
-0.5 0.5
1.05 0.5
};
\end{axis}

\end{tikzpicture}
        \caption{Bit-Flip and Phase-Flip Channels.}
        \label{fig:04_b}
    \end{subfigure}
    \caption{Fidelity as a function of the noise probability ($p=q$) between the two-qubit output state obtained via the coherent superposition of the two channels and the Bell state $\ket{\Phi^+}$. The different curves correspond to distinct choices of vacuum amplitudes, as indicated in the legend. Only specific configurations achieve unit fidelity, highlighting the critical role of the vacuum extension in enabling deterministic entanglement generation across the considered noise regimes.}
    \label{fig:04}
   \rule{\linewidth}{\arrayrulewidth}
\end{figure*}
\begin{figure*}
    \centering
    \begin{minipage}[t]{0.31\textwidth}
    \centering
    \includegraphics[width=0.95\linewidth, height=0.75\linewidth]{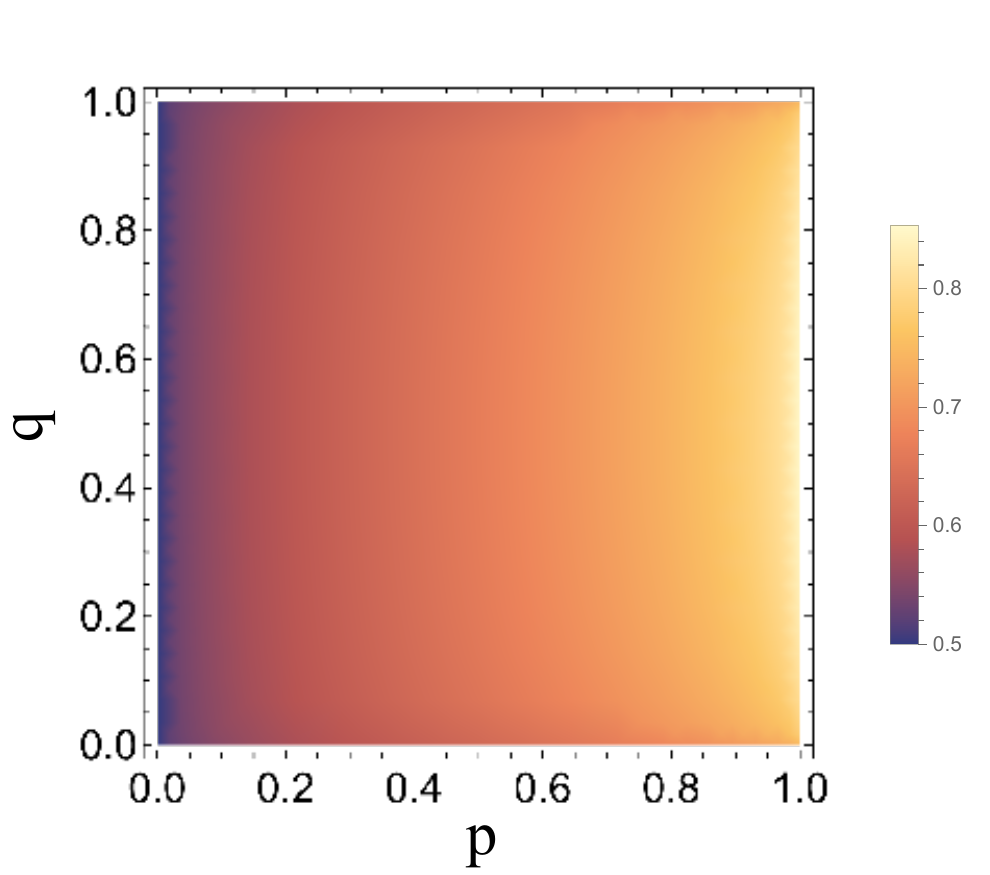}
    \caption*{(a)}
    \end{minipage}
    \hfill
    \begin{minipage}[t]{0.31\textwidth}
    \centering
    \includegraphics[width=0.95\linewidth, height=0.67\linewidth]{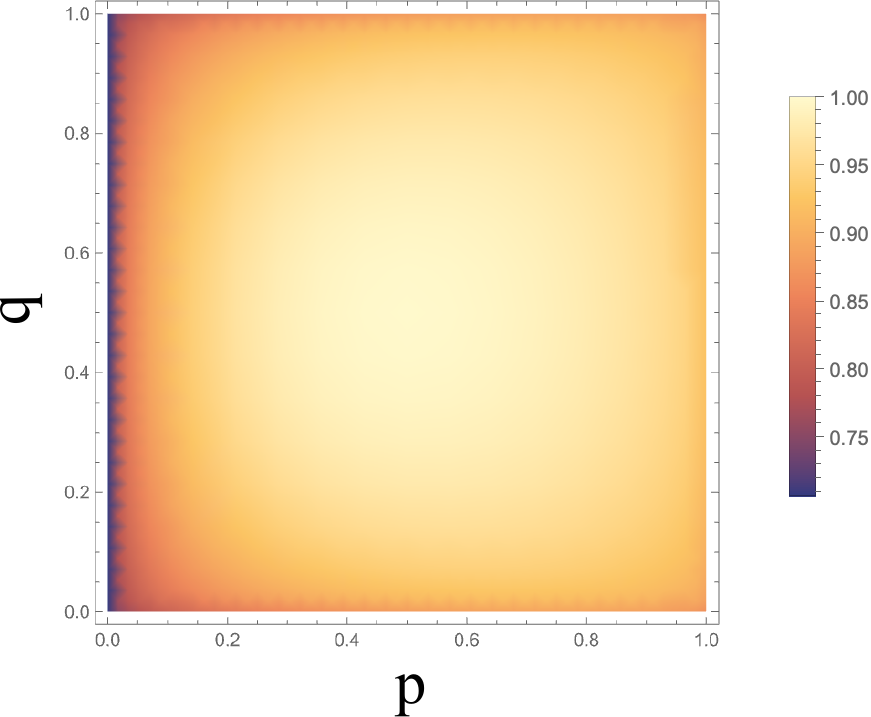}
    \caption*{(b)}
    \end{minipage}
    \hfill
    \begin{minipage}[t]{0.31\textwidth}
    \centering
    \includegraphics[width=0.9\linewidth, height=0.67\linewidth]{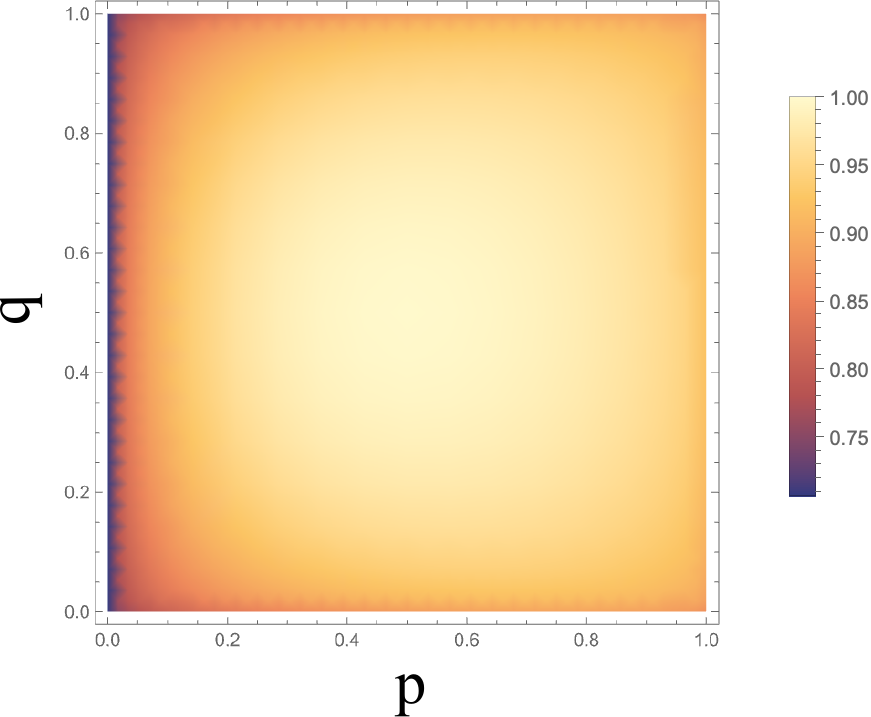}
    \caption*{(c)}
    \end{minipage}
    \caption{Density plots of the fidelity $\mathrm{fid}$ as a function of the noise probabilities $p$ and $q$. Panels (a) and (b) correspond to the bipartite case, showing the fidelity with the Bell state $\ket{\Phi^+}$ for different vacuum configurations under the superposition of bit-flip and phase-flip channels. Panel (c) corresponds to the multipartite case, showing the fidelity with the GHZ state under the same channel superposition. As predicted by Cor.~\ref{cor:01} and Cor.~\ref{cor:02}, appropriate choices of the vacuum amplitudes enable unit fidelity in the relevant operating regimes.}
    \label{fig:05}
    \rule{\linewidth}{\arrayrulewidth}
    \end{figure*}%

\begin{prop}
\label{prop:04}
Consider a two-qubit separable state $\rho_t=\ket{00}\bra{00}$ evolving under a coherent superposition of two identical depolarizing channels with noise parameters $p=q$.  
Then, there exist choices of vacuum amplitudes such that the output state is deterministically a Bell state, even in highly noisy regimes. Specifically:
\begin{itemize}
    \item For $p=q=1$ (fully depolarizing, entanglement-breaking regime), this is achieved for
\begin{equation}\label{eq:10}
    \alpha_0=\beta_0=0, \,
    \alpha_1=-\alpha_2=-\alpha_3, \, 
    \beta_1=-\beta_2=-\beta_3,
\end{equation}
with $\alpha_1=\beta_2=1/\sqrt{3}$.
    \item For $p=q=1/2$ (zero-capacity regime), this is achieved for
 \begin{equation}\label{eq:11}
    \alpha_0=-\beta_0=-\frac{1}{\sqrt{2}}, \,
    \alpha_1=-\alpha_2=-\alpha_3, \,
    \beta_1=-\beta_2=-\beta_3,
  \end{equation}
with $\alpha_1=\beta_2=1/\sqrt{6}$.
\end{itemize}
\begin{proof}
    Please refer to Appendix~\ref{Appendix:D}.
\end{proof}
\end{prop}

This result highlights that, despite each channel being individually highly noisy, their coherent superposition can yield deterministic maximal entanglement when the underlying vacuum extension is suitably chosen. This behavior arises from interference effects among the superposed spatial links, which have no classical counterpart. It also confirms that quantum noise can be harnessed as a constructive resource for entanglement generation. 

The results of Prop.~\ref{prop:04} are illustrated in Fig.~\ref{fig:04_a}, where the fidelity $\mathrm{fid}$ between the output state and the Bell state $\ket{\Phi^+}$ is plotted as a function of the noise strength $p$. The plots show that a Bell state can be deterministically generated even when the individual depolarizing channels operate in the zero-capacity ($p=q=1/2$, red curve) or entanglement-breaking regimes ($p=q=1$, blue curve).

\begin{remark}
\label{rem:02}
Our optimization is performed at the level of the channel dilation, i.e., over the space of admissible vacuum extensions. Therefore, it should be interpreted as an implementation-independent benchmark. The resulting optimal coefficients identify those coherent realizations that maximize entanglement generation under coherent superposition, without presupposing a specific hardware platform. While the exact physical realizability of such configurations depends on the details of the experimental implementation and associated constraints, the obtained solutions define target configurations within the space of admissible vacuum extensions.  As such, they provide both a fundamental performance benchmark and a guideline for the design of experimentally relevant coherent channel superpositions.
\end{remark}

The result in Prop.~\ref{prop:04} can be specialized to simpler Pauli-noise models, while preserving the same operating regimes. In particular, when the channels involve only a subset of Pauli operations, the structure of the superposition simplifies, leading to the following corollary.

\begin{cor}
\label{cor:01}
Consider a two-qubit separable state $\rho_t=\ket{00}\bra{00}$ evolving under a coherent superposition of a bit-flip channel and a phase-flip channel with noise parameters $p=q$. Then, there exist choices of vacuum amplitudes such that the output state is deterministically a maximally entangled Bell state, even in highly noisy regimes. Specifically:
\begin{itemize}
    \item For $p=q=1$, this is achieved for
\begin{equation}
\label{eq:12}
   \alpha_0=\beta_0=0, \quad \alpha_1=\beta_3=1.
\end{equation}
    \item For $p=q=1/2$, this is achieved for
 \begin{equation}
 \label{eq:13}
   \alpha_0=-\frac{1}{\sqrt{2}}, \, \alpha_1=\beta_0=\beta_3=\frac{1}{\sqrt{2}}.
  \end{equation}
\end{itemize}
\begin{proof}
    \textit{Please refer to Appendix} \ref{Appendix:D}
\end{proof}
\end{cor}
The result is Cor.~\ref{cor:01} are illustrated in \cref{fig:04_b} and \cref{fig:05}. As shown in \cref{fig:04_b} and \cref{fig:05}a and \cref{fig:05}b, if the vacuum amplitudes are not properly chosen, the output state fails to be maximally entangled, even within the same operational regime.

Moreover, in order to witness the generation of the entanglement via the spatial superposition framework, we evaluate the concurrence of the output state $\rho$ resulting from the coherent superposition of the quantum channels. The concurrence is defined as:
\begin{equation}
\label{eq:14}
    C(\rho)=max\{0,\lambda_1-\lambda_2-\lambda_3-\lambda_4\},
\end{equation} 
where $\lambda_i$ are the square roots of the  eigenvalues of the matrix $\rho\tilde{\rho}$ in decreasing order, and $\tilde{\rho}=(Y\otimes Y)\rho^*(Y\otimes Y)$ denotes the spin-flipped state, as introduced in \cite{Woo-98}.

\begin{figure*}[t]
    \centering
    \begin{subfigure}[t]{0.48\textwidth}
        \centering
        \vspace{0pt}
        \begin{tikzpicture}[spy using outlines={rectangle, magnification=2, size=1.2cm, connect spies}]

\definecolor{darkgray176}{RGB}{176,176,176}
\definecolor{dimgray}{RGB}{105,105,105}
\definecolor{gray}{RGB}{128,128,128}
\definecolor{lightgray204}{RGB}{204,204,204}
\definecolor{navy}{RGB}{0,0,128}
\definecolor{blue}{RGB}{0,0,255}
\definecolor{green}{RGB}{0,180,0}

\begin{axis}[
    width=8cm, height=6cm,
    legend cell align={left},
    legend style={
      at={(0.5,-0.30)},
      anchor=north,
      legend columns=1,
      draw=black,
      fill=white
    },
    tick align=outside,
    tick pos=left,
    grid=both,
    x grid style={lightgray204},
    xlabel={\Large{Probability $p$}},
    xmin=0.0, xmax=1.0,
    xtick style={color=black},
    xtick={0, 0.2, 0.5, 0.8, 1},
    ylabel={\Large{Concurrence}},
    ymin=0.0, ymax=1.0,
    ytick style={color=black},
    ytick={0.0, 0.4, 0.8, 1.0},
    line width=0.65pt
]

\addplot [thick, mark=*, red]
table {
0. 0.
0.1 0.0470588
0.2 0.1
0.3 0.16 
0.4 0.228571
0.5 0.307692
0.6 0.4
0.7 0.509091
0.8 0.64
0.9 0.8
1. 1.
};

\addlegendentry{\shortstack{
$\alpha_0=\beta_0=0$\\
$\alpha_1=\beta_2=\beta_3=-\alpha_2=-\alpha_3=-\beta_1=\frac{1}{\sqrt{3}}$
}}

\addplot [thick, mark=diamond, blue]
table {
0. 0.
0.05 0.35671839026838265
0.1 0.5263157894736842
0.15 0.655019811185846
0.2 0.7567567567567562
0.25 0.837100272798067 
0.3 0.8991519590239178
0.35 0.9450070102467583
0.4 0.9762633953195881
0.45 0.994227402625871
0.49999999999999806  1.
0.500000000000002 1. 
0.55 0.994508477288882 
0.6 0.9785078290616874
0.65 0.9525591057186993
0.7 0.9169805513980891
0.75 0.8717526973154062
0.8 0.8163265306122446
0.85 0.7491887410704068
0.9 0.6666666666666666
0.95 0.5582278954050992
1. 0.3076923076923076
};

\addlegendentry{\shortstack{
$\alpha_0=-1/\sqrt{2}, \beta_0=1/\sqrt{2}$\\
$\alpha_1=\beta_2=\beta_3=-\alpha_2=-\alpha_3=-\beta_1=\frac{1}{\sqrt{6}}$
}}

\end{axis}
\end{tikzpicture}
        \caption{Depolarizing Channels}
        \label{fig:06_a}
    \end{subfigure}
     \hfill
    \begin{subfigure}[t]{0.48\textwidth}
        \centering
        \vspace{0pt}
        \begin{tikzpicture}[spy using outlines={rectangle, magnification=2, size=1.2cm, connect spies}]

\definecolor{darkgray176}{RGB}{176,176,176}
\definecolor{dimgray}{RGB}{105,105,105}
\definecolor{gray}{RGB}{128,128,128}
\definecolor{lightgray204}{RGB}{204,204,204}
\definecolor{navy}{RGB}{0,0,128}
\definecolor{blue}{RGB}{0,0,255}
\definecolor{green}{RGB}{0,180,0}

\begin{axis}[
    width=8cm, height=6cm,
    legend cell align={left},
    legend style={
      fill opacity=0.8,
      draw opacity=1,
      text opacity=1,
      at={(0.5,-0.29)},     
        anchor=north,         
        legend columns=1,     
        draw=none,            
        legend cell align=left,
        align=left,
        draw=black,          
        fill=white, 
      /tikz/every even column/.append style={column sep=0.1cm}
    },
    tick align=outside,
    tick pos=left,
    grid=major,
    x grid style={lightgray204},
    xlabel={\Large{Probability $p$}},
    xmin=0.0, xmax=1.0,
    xtick style={color=black},
    xtick={0,0.2,0.5,0.8,1},
    xticklabels={
      \(\displaystyle {\ensuremath{0.0}}\),
      \(\displaystyle {\ensuremath{0.2}}\),
      \(\displaystyle {\ensuremath{0.5}}\),
      \(\displaystyle {0.8}\),
      \(\displaystyle {1}\)
    },
    ymode=log,
    log basis y={10},
    y grid style={lightgray204},
    ylabel={\Large{Concurrence}},
    ymin=0.0005, ymax=1.,
    ytick style={color=black},
    ytick={0.001,0.01, 0.1, 1},
    line width=0.65pt,
    yticklabels={
      \(\displaystyle {0.001}\),
      \(\displaystyle {0.01}\),
      \(\displaystyle {0.1}\),
      \(\displaystyle {1}\)
    }
    ]
    
    \addplot [thick,mark=*, red]
    table {%
0. 0.00001
0.02 0.019999999999999962
0.06 0.05999999999999989
0.1 0.10000000000000009
0.14 0.14000000000000018
0.18 0.18000000000000005
0.22 0.22000000000000008
0.26 0.26
0.31 0.3099999999999999
0.35 0.34999999999999987
0.39 0.3900000000000001
0.43 0.42999999999999977
0.47 0.47
0.51 0.5099999999999998
0.55 0.5499999999999998
0.59 0.5900000000000003
0.63 0.6299999999999992
0.67 0.6699999999999997
0.71 0.7100000000000006
0.75 0.7500000000000001
0.79 0.7899999999999998
0.83 0.8299999999999993
0.87 0.8699999999999999
0.91 0.9099999999999998
0.95 0.9500000000000006
0.99 0.9900000000000013
0.9999999999999998 1.
};
\addlegendentry{$\alpha_0=\beta_0=0$ and $\alpha_1=\beta_1=1$}
    \addplot [thick,mark=square, green]
    table {%
0. 0.00001
0.03 0.06386933620093954
0.07 0.10208226331936188
0.11 0.13203381603331715
0.15 0.1581104237465023
0.19 0.1818382850981494
0.23 0.20396949904680484
0.27 0.224942840998486
0.29 0.23508786056419262
0.33 0.25482755142007874
0.37 0.27396656639836997
0.41 0.29262883998416556
0.45 0.31090916210422836
0.49 0.3288813559545072
0.53 0.3466031692411051
0.57 0.3641190321760137
0.61 0.38146119724451255
0.67 0.40718789346399237
0.71 0.4241483153600154
0.75 0.4409269851976058
0.79 0.457457864580862
0.83 0.4736082909108563
0.87 0.4891139579918095
0.91 0.5034049921379314
0.95 0.5149794083982517
0.99 0.5164728099312386
1. 0.4999999999999992
};
\addlegendentry{$\alpha_0=\beta_0=\alpha_1=\beta_1=\frac{1}{\sqrt{2}}$}
    \addplot [thick,mark=diamond, blue]
    table {%
0. 0.00001
0.02 0.181818
0.04 0.279559
0.06 0.36168 
0.08 0.434389
0.1 0.5
0.14 0.614097 
0.18 0.708945
0.22 0.787147 
0.26 0.850576
0.3 0.900819
0.34 0.939302
0.38 0.96733 
0.42 0.986094 
0.46 0.996667 
0.5 1.
0.53 0.998252
0.57 0.990838
0.61 0.978169
0.65 0.960733
0.69 0.93889
0.73 0.912848
0.77 0.882642 
0.81 0.848065
0.85 0.808526
0.89 0.762718 
0.93 0.707613 
0.97 0.633866 
0.99 0.576302
};
\addlegendentry{$\alpha_0=-\frac{1}{\sqrt{2}}$ and $\beta_0=\alpha_1=\beta_1=\frac{1}{\sqrt{2}}$}
\end{axis}

\end{tikzpicture}
        \caption{Bit-Flip and Phase-Flip Channels}
        \label{fig:06_b}
    \end{subfigure}
    \caption{Concurrence versus noise probability ($p=q$) for the two-qubit output state in the Bell-state generation framework. The curves correspond to different vacuum configurations. For the optimal amplitudes, unit concurrence is achieved at $p=q=1/2$ (blue curve) and $p=q=1$ (red curve), proving Bell-state generation even in the zero-capacity and fully depolarizing regimes.}
    \label{fig:06}
    \rule{\linewidth}{\arrayrulewidth}
\end{figure*}
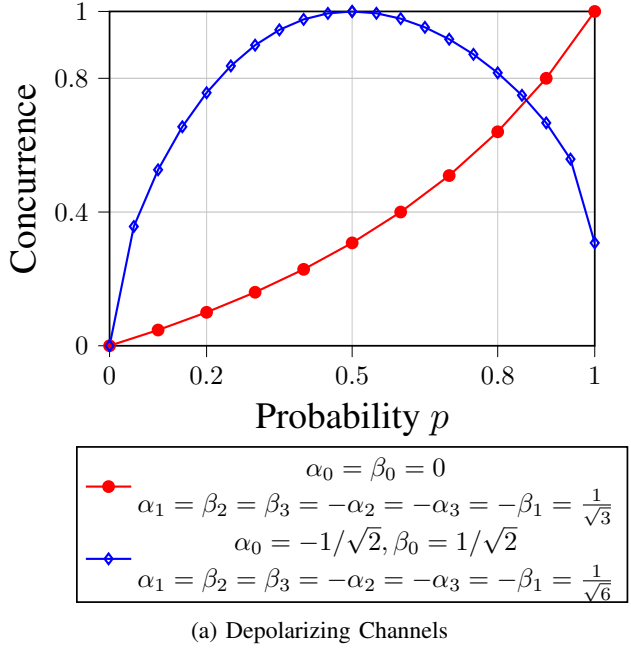
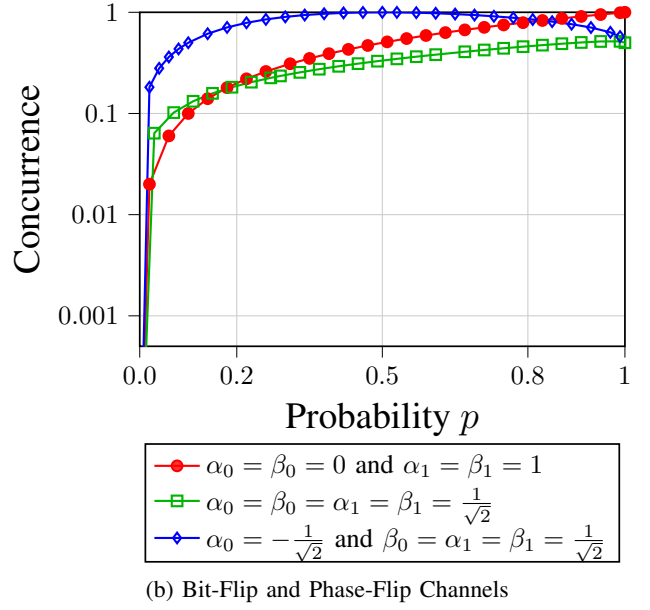

Fig.~\ref{fig:06} reports the concurrence as a function of the noise probability in the symmetric case $p=q$, for the depolarizing channels and for the superposition of bit- and phase-flip channels, respectively. Their behavior was previously analyzed in terms of fidelity in Fig.~\ref{fig:04}. In full analogy with the fidelity results, the concurrence attains its maximal value, a sign of generation of a maximally entangled state, for suitable choices of the vacuum amplitudes, as derived by the theoretical analysis in Prop.~\ref{prop:04} and Cor.~\ref{cor:01}. Notice that for the depolarizing channels case, in \cref{fig:06_a}, the case of vacuum amplitudes all equal to $1/\sqrt{4}$ is not reported because as evident from the fidelity, the state remains separable and so the concurrence is trivially zero.  Remarkably, this occurs in both the considered adverse regimes.


\subsection{GHZ State Generation under Noise}
As was done for the case of Bell states, the generation of a GHZ state can be generalized to the case of noisy channels as well. Given the results concerning Bell states generated through the superposition of depolarizing channels and the superposition of bit-flip and phase-flip channels.
\begin{prop}
\label{prop:05}
Consider a n-qubit separable state $\rho_t=\ket{0}^{\otimes n}\bra{0}^{\otimes n}$ evolving under a coherent superposition of two depolarizing channels with noise parameters $p=q$.  
Then, there exist choices of vacuum amplitudes such that the output state is deterministically a maximally entangled $n$-qubit GHZ state, even in highly noisy regimes, up to local unitary transformations. Specifically:
\begin{itemize}
    \item For $p=q=1$ (fully depolarizing, entanglement-breaking regime), this is achieved for
\begin{align}\label{eq:15}
    &\alpha_0=\beta_0=0, \,
    \alpha_1=\alpha_2=-\alpha_3=\frac{1}{\sqrt{3}},\nonumber\\ 
    &\beta_1=\beta_2=-\beta_3=-\frac{1}{\sqrt{3}}.
\end{align}
    \item For $p=q=1/2$ (zero-capacity regime), this is achieved for
 \begin{align}\label{eq:16}
    &\alpha_0=-\beta_0=-\frac{1}{\sqrt{2}},\,
    \alpha_1=\alpha_2=-\alpha_3=\frac{1}{\sqrt{6}},\nonumber \\
    &\beta_1=\beta_2=-\beta_3=-\frac{1}{\sqrt{6}}.
  \end{align}
\end{itemize}
\begin{proof}
    Please refer to Appendix~\ref{Appendix_E}.
\end{proof}
\end{prop}

The results of Prop. \ref{prop:05} can be specialized to simpler Pauli-noise models, while preserving the same operating regimes. This is captured by Cor.~\ref{cor:02}, where the superposition of bit-flip and phase-flip channels is considered for the generation of GHZ states. 
\begin{cor}
\label{cor:02}
Consider a n-qubit separable state $\rho_t=\ket{0}^{\otimes n}\bra{0}^{\otimes n}$ evolving under a coherent superposition of bit-flip, $\mathcal{F}(\cdot)$ and phase-flip, $\mathcal{N}(\cdot)$ channels with noise parameters $p=q$.  
Then, there exist choices of vacuum amplitudes such that the output state is deterministically a maximally entangled GHZ state, even in highly noisy regimes. Specifically:
\begin{itemize}
    \item For $p=q=1$, this is achieved for
\begin{equation}\label{eq:17}
    \alpha_0=\beta_0=1, \,
    \alpha_1 =\beta_1 =1.
\end{equation}
    \item For $p=q=1/2$, this is achieved for
 \begin{equation}\label{eq:18}
    \alpha_0=-\frac{1}{\sqrt{2}}, \,
    \beta_0=\alpha_1 =\beta_1 =1/\sqrt{2}.
  \end{equation}
\end{itemize}
\begin{proof}
    \textit{Please refer to Appendix} \ref{Appendix_E}
\end{proof}
\end{cor}

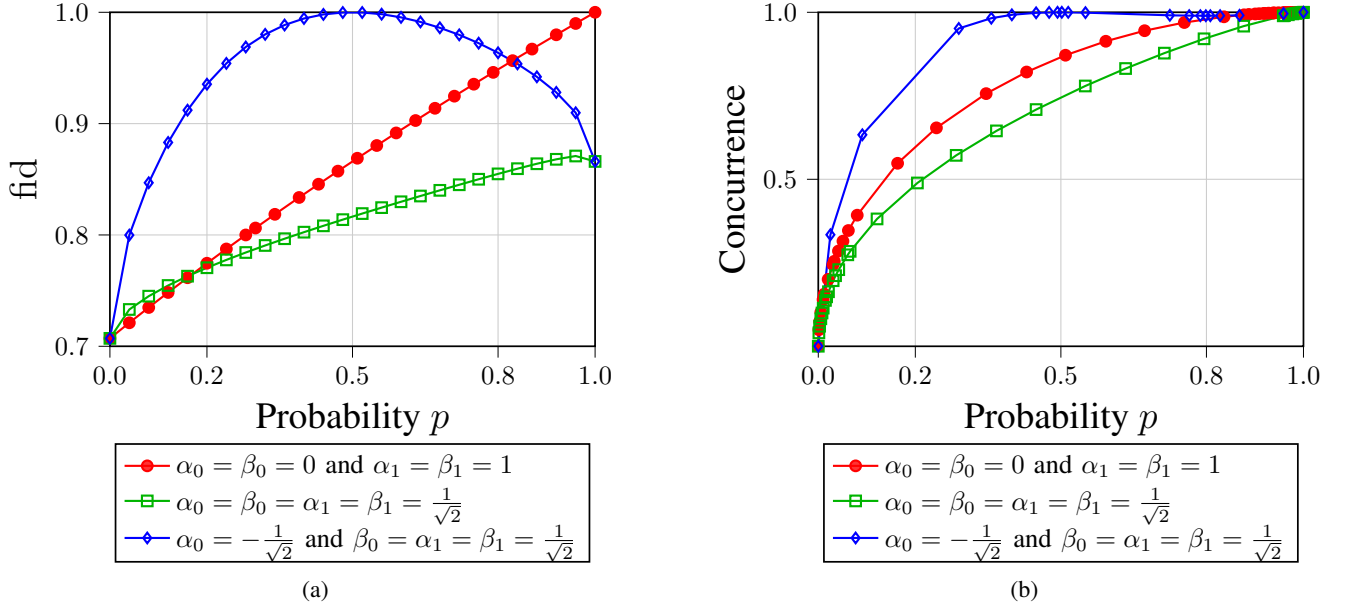
\begin{figure*}[t]
    \centering
    \begin{subfigure}[t]{0.48\textwidth}
        \centering
        \vspace{0pt}
        \begin{tikzpicture}[spy using outlines={rectangle, magnification=2, size=1.2cm, connect spies}]

\definecolor{darkgray176}{RGB}{176,176,176}
\definecolor{dimgray}{RGB}{105,105,105}
\definecolor{gray}{RGB}{128,128,128}
\definecolor{lightgray204}{RGB}{204,204,204}
\definecolor{navy}{RGB}{0,0,128}
\definecolor{blue}{RGB}{0,0,255}
\definecolor{green}{RGB}{0,180,0}

\begin{axis}[
    width=8cm, height=6cm,
    legend cell align={left},
    legend style={
      fill opacity=0.8,
      draw opacity=1,
      text opacity=1,
      at={(0.5,-0.29)},     
        anchor=north,         
        legend columns=1,     
        draw=none,            
        legend cell align=left,
        align=left,
        draw=black,          
        fill=white, 
      /tikz/every even column/.append style={column sep=0.1cm}
    },
    tick align=outside,
    tick pos=left,
    grid=major,
    x grid style={lightgray204},
    xlabel={\Large{Probability $p$}},
    xmin=0.0, xmax=1.0,
    xtick style={color=black},
    xtick={0.0,0.2,0.5,0.8,1.0},
    xticklabels={
      \(\displaystyle {\ensuremath{0.0}}\),
      \(\displaystyle {0.2}\),
      \(\displaystyle {0.5}\),
      \(\displaystyle {0.8}\),
      \(\displaystyle {1.0}\)
    },    
    y grid style={lightgray204},
    ylabel={\Large{$\mathrm{fid}$}},
    ymin=0.7, ymax=1.0,
    ytick style={color=black},
    ytick={0.7, 0.8,0.9, 1},
    line width=0.65pt,
    yticklabels={
      \(\displaystyle {0.7}\),
      \(\displaystyle {0.8}\),
      \(\displaystyle {0.9}\),
      \(\displaystyle {1.0}\),
    }
    ]
    
    \addplot [thick,mark=*, red]
    table {%
0. 0.7071067811865476
0.04 0.7211102550927979
0.08 0.7348469228349535
0.12 0.7483314773547883
0.16 0.7615773105863908
0.2 0.7745966692414834
0.24 0.7874007874011811
0.28 0.8
0.3 0.8062257748298549
0.34 0.818535277187245
0.39 0.8336666000266534
0.43 0.8455767262643882
0.47 0.8573214099741123
0.51 0.8689073598491384
0.55 0.8803408430829505
0.59 0.8916277250063503
0.63 0.9027735042633894
0.67 0.9137833441248533
0.71 0.9246621004453465
0.75 0.9354143466934853
0.79 0.9460443964212251
0.83 0.9565563234854495
0.87 0.9669539802906858
0.92 0.9797958971132712
0.96 0.9899494936611666
1. 1.
};
\addlegendentry{$\alpha_0=\beta_0=0$ and $\alpha_1=\beta_1=1$}
    \addplot [thick, mark=square, green]
    table {%
0. 0.7071067811865476
0.04 0.7330641967194645
0.08 0.7450096090045889
0.12 0.754597647365743
0.16 0.7629630943823944
0.2 0.770551750371122
0.24 0.7775951246961033
0.28 0.784232348172637
0.32 0.7905550545236641
0.36 0.7966275068156915
0.4 0.8024968458753736
0.44 0.8081987839530858
0.48 0.8137609580891234
0.52 0.8192049644868443
0.56 0.8245475742377298
0.6 0.8298013724376765
0.64 0.8349749089592748
0.68 0.8400723230520526
0.72 0.8450922447033372
0.76 0.8500254887100992
0.8 0.8548504142651103
0.84 0.8595231313900822
0.88 0.8639544424483675
0.92 0.8679442567295498
0.96 0.8709149830298863
1. 0.8660254037844386
};
\addlegendentry{$\alpha_0=\beta_0=\alpha_1=\beta_1=\frac{1}{\sqrt{2}}$}
    \addplot [thick,mark=diamond,blue]
    table {%
0. 0.707107
0.04 0.799862
0.08 0.846873 
0.12 0.883084 
0.16 0.912069
0.2 0.935414
0.24 0.954098 
0.28 0.968828 
0.32 0.980165 
0.36 0.988571
0.4 0.994435 
0.44 0.998084
0.48 0.999796 
0.52 0.999804
0.56 0.9983 
0.6 0.99544 
0.64 0.991342 
0.68 0.986091 
0.72 0.979731 
0.76 0.972263 
0.8 0.963624 
0.84 0.953659 
0.88 0.942039 
0.92 0.928052
0.96 0.90975 
1. 0.866025
};
 \addlegendentry{$\alpha_0=-\frac{1}{\sqrt{2}}$ and $\beta_0=\alpha_1=\beta_1=\frac{1}{\sqrt{2}}$}
\addplot [thick, black!30!red, dotted]
table {%
    -0.5 0.5
    1.05 0.5
};
\end{axis}

\end{tikzpicture}
        \caption{}
        \label{fig:07_a}
    \end{subfigure}
     \hfill
    \begin{subfigure}[t]{0.48\textwidth}
        \centering
        \vspace{0pt}
        \begin{tikzpicture}[spy using outlines={rectangle, magnification=2, size=1.2cm, connect spies}]

\definecolor{darkgray176}{RGB}{176,176,176}
\definecolor{dimgray}{RGB}{105,105,105}
\definecolor{gray}{RGB}{128,128,128}
\definecolor{lightgray204}{RGB}{204,204,204}
\definecolor{navy}{RGB}{0,0,128}
\definecolor{blue}{RGB}{0,0,255}
\definecolor{green}{RGB}{0,180,0}

\begin{axis}[
    width=8cm, height=6cm,
    legend cell align={left},
    legend style={
      fill opacity=0.8,
      draw opacity=1,
      text opacity=1,
      at={(0.5,-0.29)},     
        anchor=north,         
        legend columns=1,     
        draw=none,            
        legend cell align=left,
        align=left,
        draw=black,          
        fill=white, 
      /tikz/every even column/.append style={column sep=0.1cm}
    },
    tick align=outside,
    tick pos=left,
    grid=major,
    x grid style={lightgray204},
    xlabel={\Large{Probability $p$}},
    xmin=0.0, xmax=1.,
    xtick style={color=black},
    xtick={0.0,0.2,0.5,0.8,1.0},
    xticklabels={
      \(\displaystyle {\ensuremath{0.0}}\),
      \(\displaystyle {0.2}\),
      \(\displaystyle {0.5}\),
      \(\displaystyle {0.8}\),
      \(\displaystyle {1.0}\)
    },
    y grid style={lightgray204},
    ylabel={\Large{Concurrence}},
    ymin=0.0, ymax=1.,
    ytick style={color=black},
    ytick={0.5, 1},
    line width=0.65pt,
    yticklabels={
      \(\displaystyle {0.5}\),
      \(\displaystyle {1.0}\),
    }
    ]
    
    \addplot [thick,mark=*, red]
    table {%
0.0 0.00020203050728717973
0.0012268104577487112 0.04951884339923532
0.002760298019730519 0.07424942285501741
0.005213878118901411 0.10198319328577836
0.009814340804846833 0.13975822095411808
0.012267920904017723 0.15615806071006397
0.020293651697903155 0.20043819769836854
0.03026850914349452 0.24417378164172154
0.03292847112898555 0.25450473089331493
0.04152927978810532 0.2851909509372481
0.05084312032780537 0.3148034589564466
0.06199536889819733 0.346622722901386
0.08025764881887605 0.3925226202957201
0.16346277092346467 0.5479100878510563
0.24349293774233727 0.6539855233522225
0.34598550825253294 0.756482018678767
0.429331209352014 0.8211803281742477
0.5095019551657791 0.8714422918437035
0.5926201946734201 0.9132588319923755
0.6725634788953451 0.944873179134047
0.7544338486478805 0.9693798354154618
0.8361904876044959 0.9864919886389079
0.8774346889493128 0.9924604498553321
0.8873891832862444 0.9936391719125514
0.900485571227249 0.9950361191766021
0.9085399183341507 0.9958087434149573
0.9178333957651912 0.9966186076672061
0.9202629362589725 0.9968159311858732
0.9281589428637615 0.9974160929669962
0.9372697197154413 0.9980305165351512
0.9399056420955427 0.9981927009090233
0.9484723898308721 0.9986715703324384
0.9583570987562524 0.9991325581603292
0.9608167499614495 0.9992320415781394
0.964506226769245 0.9993698975163517
0.9688106163783399 0.9995134928300373
0.9734224623880844 0.9996467548561777
0.9780343083978288 0.999758725089428
0.9807800172970798 0.999815279071539
0.9835257261963308 0.999864289942711
0.9862714350955818 0.9999057588121318
0.9890171439948328 0.9999396866181328
0.9907332120568646 0.9999570623987897
0.9924492801188964 0.9999714929083116
0.9945085617933347 0.9999849219395374
0.9958814162429602 0.999991518597951
0.9972542706925858 0.9999962304781805
0.9986271251422112 0.9999990576068684
0.9999999795918367 0.9999999999999998
};
 \addlegendentry{$\alpha_0=\beta_0=0$ and $\alpha_1=\beta_1=1$} 
    \addplot [thick, mark=square, green]
    table {%
0.0 0.0001649532921231538
0.0012268104577487112 0.04021045941161094
0.002760298019730519 0.06014262627304889
0.005213878118901411 0.08239443455877696
0.007667458218072303 0.09967127844113047
0.010121038317243194 0.11427542398353917
0.014721501003188616 0.13737051869986525
0.017175081102359508 0.14815145678343142
0.020958642194275914 0.16330993516169667
0.0306010043916809 0.19642403077125015
0.035588433114476584 0.21138985695423063
0.042150202490751985 0.22948428679425478
0.06077788357015209 0.2739195812884868
0.06564782488233308 0.2842903231615861
0.12110912009821138 0.38135588646067653
0.20425737680483996 0.4889795245004558
0.2842306782257526 0.5716545867226278
0.3671514733405411 0.644908968970957
0.44893812949614426 0.7087453173281826
0.5503099851466406 0.7793828301919457
0.6333713592563216 0.8316522686532452
0.7132577780802865 0.8777164117415459
0.7950712824348619 0.9206377420464908
0.8767710559935172 0.958577588126218
0.9583570987562524 0.9894515630616605
0.9608167499614495 0.9902190979072191
0.9688106163783399 0.9926198469933389
0.9780343083978288 0.9951885979678273
0.9842121534211435 0.9967654714036184
0.9893603576072392 0.9979719881877563
0.9948517754057411 0.9991218793539063
0.9975974843049922 0.9996251257751386
0.999313552367024 0.9999028622229295
0.9999999795918367 0.9999999974471575
};
 \addlegendentry{$\alpha_0=\beta_0=\alpha_1=\beta_1=\frac{1}{\sqrt{2}}$} 
    \addplot [thick,mark=diamond,blue]
    table {%
0 0.000285735
0.0249486 0.333982 
0.0908231 0.633281
0.289559 0.951621
0.356568 0.982287
0.398949 0.992471
0.448938 0.998408
0.475778 0.999677 
0.493084 0.999975
0.501901 0.999998
0.515109 0.999893
0.550644 0.998958
0.72457 0.990974
0.764485 0.990026
0.787473 0.989779
0.798771 0.989744
0.808022 0.989758
0.828171 0.989929
0.868203 0.990852
0.958972 0.995999 
1. 1.
};
 \addlegendentry{$\alpha_0=-\frac{1}{\sqrt{2}}$ and $\beta_0=\alpha_1=\beta_1=\frac{1}{\sqrt{2}}$}
    
\end{axis}

\end{tikzpicture}
        \caption{}
        \label{fig:07_b}
    \end{subfigure}
    \caption{(a) Fidelity versus noise probability ($p=q$) of GHZ state in case of the superposition of bit-flip and phase-flip channels (b) Concurrence versus noise probability ($p=q$) of GHZ state obtained from the superposition of the bit-flip and phase-flip channel $\mathcal{N}$ for different configurations of vacuum amplitudes.}
    \label{fig:07}
    \hrulefill
\end{figure*}

The results of Cor.~\ref{cor:02} are numerically validated in \cref{fig:07_a} and \cref{fig:05}(c) in terms of fidelity. The obtained results are in full agreement with the theoretical predictions. The numerical analysis associated with Prop.~\ref{prop:05} is instead omitted due to space limitations, as it does not provide additional insights beyond the bipartite depolarizing-channel case. We therefore focus on the representative scenario of bit-flip and phase-flip channel superposition.

We further assess entanglement in \cref{fig:07_b}, by calculating the average concurrence over all qubit pairs. While this quantity reflects the distribution of pairwise correlations, it is not a genuine multipartite entanglement measure, and unit values may occur for different vacuum configurations \cite{AloVis-17}.
\subsection{W State Generation under Noise}
\begin{figure*}[t]
    \centering
    \begin{subfigure}[t]{0.45\textwidth}
        \centering
        \vspace{0pt}
        \begin{tikzpicture}[spy using outlines={rectangle, magnification=2, size=1.2cm, connect spies}]

\definecolor{darkgray176}{RGB}{176,176,176}
\definecolor{dimgray}{RGB}{105,105,105}
\definecolor{gray}{RGB}{128,128,128}
\definecolor{lightgray204}{RGB}{204,204,204}
\definecolor{navy}{RGB}{0,0,128}
\definecolor{blue}{RGB}{0,0,255}
\definecolor{green}{RGB}{0,180,0}

\begin{axis}[
    width=8cm, height=6cm,
    legend cell align={left},
    legend style={
     fill opacity=0.8,
      draw opacity=1,
      text opacity=1,
      at={(0.5,-0.29)},     
        anchor=north,         
        legend columns=1,     
        draw=none,            
        legend cell align=left,
        align=left,
        draw=black,          
        fill=white, 
      /tikz/every even column/.append style={column sep=0.1cm}
    },
    tick align=outside,
    tick pos=left,
    grid=major,
    x grid style={lightgray204},
    xlabel={\Large{Probability $p$}},
    xmin=0.0, xmax=1.,
    xtick style={color=black},
    xtick={0,0.5,1},
    xticklabels={
      \(\displaystyle {\ensuremath{0}}\),
      \(\displaystyle {0.5}\),
      \(\displaystyle {1}\)
    },
    y grid style={lightgray204},
    ylabel={\Large{$\mathrm{fid}$}},
    ymin=0.0, ymax=1.,
    ytick style={color=black},
    ytick={0, 0.5, 1},
    line width=0.65pt,
    yticklabels={
      \(\displaystyle {0.}\),
      \(\displaystyle {0.5}\),
      \(\displaystyle {1.0}\),
    }
    ]
    
    \addplot [thick,mark=*, red]
    table {%
2.040816326530612e-8 0.00014285714285714287
0.0012268104577487112 0.03502585413303595
0.002760298019730519 0.05253853842400376
0.005213878118901411 0.0722071888311781
0.009814340804846833 0.09906735488972558
0.012267920904017723 0.11076064691043352
0.020293651697903155 0.14245578857281707
0.03026850914349452 0.17397847321865575
0.03292847112898555 0.18146203770757552
0.04152927978810532 0.2037873396168303
0.05084312032780537 0.22548419086003654
0.12110912009821138 0.3480073563851939
0.20425737680483996 0.45194842272635494
0.2842306782257526 0.5331328898368142
0.3671514733405411 0.6059302545182418
0.44893812949614426 0.670028454243657
0.5716599705280151 0.7560819866443158
0.6531622996938181 0.8081845703141197
0.7345508980637009 0.8570594483836583
0.8148053574743983 0.902665695301643
0.8980073105789717 0.9476324765324223
0.9783775220102351 0.9891296790665192
0.9890171439948328 0.9944934107347484
0.9917628528940837 0.995872910011154
0.9958814162429602 0.9979385834022856
0.9986271251422112 0.9993133268110715
0.9999999795918367 0.9999999897959184
};
 \addlegendentry{$\alpha^{(i)}_0=0$ and $\alpha^{(i)}_1=1$} 
    \addplot [thick, mark=square, green]
    table {%
2.040816326530612e-8 0.0000824786103050415
0.0012268104577487112 0.020228391364521064
0.002760298019730519 0.030354093120054687
0.005213878118901411 0.04174328651906039
0.009814340804846833 0.05733741926958042
0.012267920904017723 0.06414472251816754
0.020293651697903155 0.08266736288864011
0.03026850914349452 0.10121534777106474
0.042150202490751985 0.11980212401499406
0.06077788357015209 0.14454843004544018
0.1424808261288228 0.22613403394804305
0.22407003789157337 0.2900236053675977
0.305545518858404 0.3467187885128472
0.3869072690293145 0.3998785861946057
0.47019610473083556 0.45266404790710224
0.5503099851466406 0.5030608185828315
0.6333713592563216 0.5558509111280616
0.7132577780802865 0.607899593235937
0.7950712824348619 0.6632491872764659
0.8767710559935172 0.7213787245806836
0.9583570987562524 0.7831746529039938
0.9794071628474543 0.7998522124257018
0.9893603576072392 0.8078550759870841
0.9945085617933347 0.8120250841709243
0.9972542706925858 0.8142577716985064
0.9996567659794304 0.8162163796025342
0.9999999795918367 0.8164965642645309
};
 \addlegendentry{$\alpha^{(i)}_0=\alpha^{(i)}_1=\frac{1}{\sqrt{2}}$} 
    \addplot [thick,mark=diamond, blue]
    table {%
2.040816326530612e-8 0.00009759000769317988
0.0012268104577487112 0.02393308651521805
0.002760298019730519 0.03591048463958264
0.005213878118901411 0.04937841807875037
0.009814340804846833 0.06780911845048873
0.012267920904017723 0.07585026025251597
0.020293651697903155 0.09771322988535629
0.03026850914349452 0.11957600520982528
0.042150202490751985 0.1414475918433712
0.06077788357015209 0.1704984780687762
0.1424808261288228 0.2655363555439828
0.22407003789157337 0.33891113818166946
0.305545518858404 0.4030399216748221
0.3869072690293145 0.46218682493038393
0.47019610473083556 0.5198593596403986
0.5503099851466406 0.5738446110500932
0.6333713592563216 0.6291781653207582
0.7132577780802865 0.6824419305731434
0.7950712824348619 0.7376040886821762
0.8767710559935172 0.793828355830491
0.9583570987562524 0.851618688124551
0.9794071628474543 0.8668591996816205
0.9893603576072392 0.87411829211403
0.9945085617933347 0.877886787894062
0.9972542706925858 0.8799005639004616
0.9989703387546176 0.8811605684065983
0.9999999795918367 0.8819170886896069
};
 \addlegendentry{$\alpha^{(i)}_0=\frac{1}{\sqrt{3}}$ and $\alpha^{(i)}_1=\sqrt{\frac{2}{3}}$} 
\addplot [thick, black!30!red, dotted]
table {%
    -0.5 0.468
    1.05 0.468
};
     \addlegendentry{$F=0.468$} 
\end{axis}

\end{tikzpicture}
        \caption{}
        \label{fig:08_a}
    \end{subfigure}
     \hfill
    \begin{subfigure}[t]{0.45\textwidth}
        \centering
        \vspace{0pt}
        \begin{tikzpicture}[spy using outlines={rectangle, magnification=2, size=1.2cm, connect spies}]

\definecolor{darkgray176}{RGB}{176,176,176}
\definecolor{dimgray}{RGB}{105,105,105}
\definecolor{gray}{RGB}{128,128,128}
\definecolor{lightgray204}{RGB}{204,204,204}
\definecolor{navy}{RGB}{0,0,128}

\begin{axis}[
    width=8cm, height=6cm,
    legend cell align={left},
    legend style={
   fill opacity=0.8,
      draw opacity=1,
      text opacity=1,
      at={(0.5,-0.29)},     
        anchor=north,         
        legend columns=1,     
        draw=none,            
        legend cell align=left,
        align=left,
        draw=black,          
        fill=white, 
      /tikz/every even column/.append style={column sep=0.1cm}
    },
    tick align=outside,
    tick pos=left,
    grid=major,
    x grid style={lightgray204},
    xlabel={\Large{Probability $p$}},
    xmin=-0.0, xmax=1.0,
    xtick style={color=black},
    xtick={0,0.5,1},
    xticklabels={
      \(\displaystyle {\ensuremath{0}}\),
      \(\displaystyle {0.5}\),
      \(\displaystyle {1}\)
    },
    y grid style={lightgray204},
    ylabel={\Large{Concurrence}},
    ymin=0.0, ymax=1.0,
    ytick style={color=black},
    ytick={0.0, 0.5, 1.0},
    line width=0.65pt,
    yticklabels={
      \(\displaystyle {0.3}\),
      \(\displaystyle {0.5}\),
      \(\displaystyle {1.0}\),
    }
    ]
    
    \addplot [thick,mark=*, red]
    table {%
2.040816326530612e-8 0.00016495721889176756
0.0012268104577487112 0.040436102182440564
0.002760298019730519 0.06063836267956854
0.005213878118901411 0.08330519480044399
0.009814340804846833 0.11420585921748105
0.012267920904017723 0.12763360899899945
0.020293651697903155 0.16393646844565563
0.03026850914349452 0.1998770119456388
0.03292847112898555 0.20838119895578433
0.04152927978810532 0.23367894241135115
0.05084312032780537 0.2581509791650766
0.06199536889819733 0.2845211813476316
0.08025764881887605 0.32271876732446325
0.16346277092346467 0.45395459160699514
0.24349293774233727 0.5461745762814983
0.32647059825508584 0.6228353601751774
0.408314119808649 0.6857994326140143
0.4900439105662921 0.7393657033538263
0.5716599705280151 0.7854750267421201
0.6531622996938181 0.8253930805891857
0.7345508980637009 0.8599968145535098
0.8148053574743983 0.8895712579924762
0.8980073105789717 0.9159342047897537
0.9780343083978288 0.9375030232342999
0.9807800172970798 0.9381804160920526
0.9903899984444582 0.9405193946407274
0.9951949890181475 0.9416703600842284
0.9979406979173985 0.9423225343544253
0.9996567659794304 0.9427281093059033
0.9999999795918367 0.9428090367718132
};
 \addlegendentry{$\alpha^{(i)}_0=0$ and $\alpha^{(i)}_1=1$} 
    \addplot [thick, mark=square, green]
    table {%
2.040816326530612e-8 0.00011664237347767758
0.0012268104577487112 0.028604338826886015
0.002760298019730519 0.04291728106276053
0.005213878118901411 0.05900819964811163
0.009814340804846833 0.08102068326696384
0.012267920904017723 0.09062097643369374
0.020293651697903155 0.11670939843018072
0.03026850914349452 0.14277304547180675
0.03292847112898555 0.14898070050396395
0.04152927978810532 0.16755126484704067
0.05084312032780537 0.18568026232017146
0.12110912009821138 0.2899950281994531
0.20425737680483996 0.3819819954697029
0.2842306782257526 0.4568155766167293
0.3671514733405411 0.5266249096775286
0.44893812949614426 0.5904960568870079
0.5306110548558273 0.6507874163428498
0.6121702494195903 0.7082775889888551
0.6936157131874332 0.763305300783067
0.7739270379960906 0.8151969823465072
0.857185856498624 0.8660507461461713
0.9372697197154413 0.9111672830603816
0.9794071628474543 0.9328610338828551
0.9893603576072392 0.9377280846201945
0.9945085617933347 0.9402027668707105
0.9972542706925858 0.9415102730214129
0.999313552367024 0.9424851697629322
0.9999999795918367 0.9428090319615627
};
 \addlegendentry{$\alpha^{(i)}_0=\alpha^{(i)}_1=\frac{1}{\sqrt{2}}$} 
    \addplot [thick,mark=diamond,blue]
    table {%
2.040816326530612e-8 0.00012777531197334668
0.0012268104577487112 0.031331919449097974
0.002760298019730519 0.04700486670038195
0.005213878118901411 0.064617737503959
0.009814340804846833 0.08869545540871418
0.012267920904017723 0.0991887919018542
0.020293651697903155 0.1276746259557197
0.03026850914349452 0.15608130740026518
0.03292847112898555 0.1628380528959208
0.04152927978810532 0.1830281236888143
0.05084312032780537 0.2027015581576248
0.12110912009821138 0.31500476292731894
0.20425737680483996 0.41234503871226824
0.2842306782257526 0.4899977361137894
0.3869072690293145 0.576779318508714
0.47019610473083556 0.6400251912711483
0.5503099851466406 0.6963061713924465
0.6333713592563216 0.7506614980732019
0.7132577780802865 0.7993952872404102
0.7950712824348619 0.8456972060308724
0.857185856498624 0.8782395103939165
0.9372697197154413 0.9164122034328797
0.9787207356226415 0.9342395088543048
0.9890171439948328 0.9384379646830855
0.9945085617933347 0.9406375688687307
0.9972542706925858 0.941726848953641
0.999313552367024 0.9425391559726808
0.9999999795918367 0.9428090335649794
};
 \addlegendentry{$\alpha^{(i)}_0=\frac{1}{\sqrt{3}}$ and $\alpha^{(i)}_1=\sqrt{\frac{2}{3}}$} 
    
\end{axis}

\end{tikzpicture}
        \caption{}
        \label{fig:08_b}
    \end{subfigure}
    \caption{(a) Fidelity versus noise probability ($p=q$) between the 3-qubit state obtained from the superposition of the three channels and a 3-qubit W state. (b) Concurrence versus noise probability ($p_i=p$ $\forall i$) between the 3-qubit W state obtained from the superposition of the three channels $\mathcal{F}_i(\cdot)$ for three different configurations of vacuum amplitudes.}
    \label{fig:08}
    \hrulefill
\end{figure*}
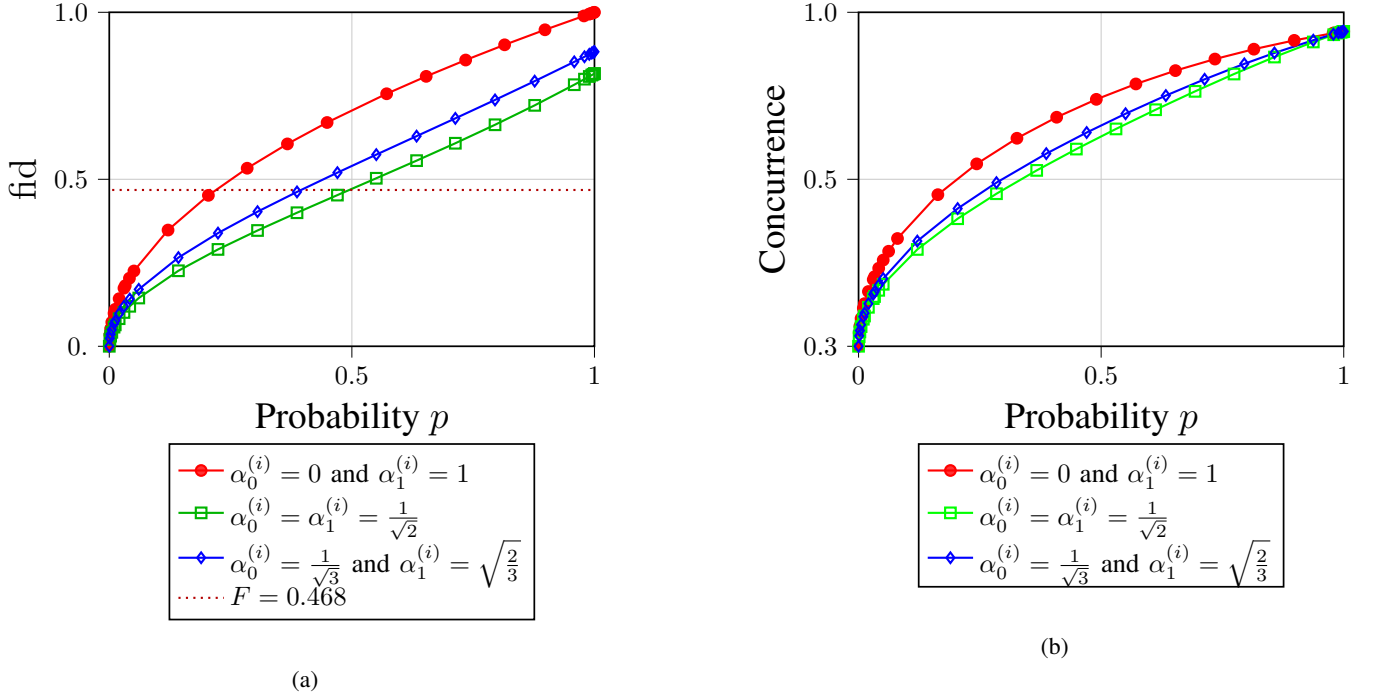

 The W-state case is more subtle. While GHZ states can be deterministically generated even by exploiting fully depolarizing noise as a constructive resource, extending this mechanism to W states appears to be less straightforward. In particular, our analysis suggests that their generation requires a more structured noise model, such as the generalized memoryless bit-flip channel defined as follows:
\begin{equation}
\label{eq:19}
    \mathcal{F}_i(\rho)=(1-p_i)\rho+p_i X_i\rho X_i,
\end{equation}
with $X_i$ given in Sec.~\ref{sec:3.2}. This asymmetry can be traced back to the distinct entanglement structures of GHZ and W states. GHZ states exhibit an entanglement structure, which can be generated through the spatial superposition framework by harnessing interference effects activated by a proper tuning of the vacuum coefficients, even under uniformly mixing noise. In contrast, W states rely on a different correlation structure, which does not appear to be generated via interference mechanisms in presence of depolarizing dynamics. A deeper investigation of the reasons behind this asymmetry is left for future work.
\begin{prop}
\label{prop:07}
An $n$-qubit separable state $\rho_t=\ket{0}^{\otimes n}\bra{0}^{\otimes n}$, with the control qudit prepared in the state $\ket{\tilde{0}}\bra{\tilde{0}}$, evolving under a coherent superposition of $n$ noisy channels $\{\mathcal{F}_i(\cdot)\}_{i=0}^{n-1}$, deterministically yields an $n$-qubit W state in the asymptotic limit $p_i=1$, for vacuum amplitudes $\alpha^{(i)}_0=0$ and $\alpha^{(i)}_1=1$.
\begin{proof}
Please refer to Appendix~\ref{Appendix_F}.
\end{proof}
\end{prop}
In this case, no parameter regime achieving unit fidelity in the zero-capacity limit is observed. However, as shown in \cref{fig:08_a}, where the fidelity is plotted as a function of the noise probability (for $p_i=p$), the obtained values remain above the threshold required for entanglement purification \cite{MigRieDur-23}.

Also for the W state case, the degree of entanglement is assessed via the average concurrence. We recall that W states belong to a different class of multipartite entanglement compared to GHZ states. Accordingly their average concurrence is strictly less than one. This reflects the fact that average concurrence quantifies pairwise correlations. As shown in \cref{fig:08_b}, for the case $n=3$ and $p_0=p_1=p_2=p$, the concurrence reaches the expected value for a $3$-qubit W state, namely $C=2\sqrt{2}/3$.
\section{Conclusions}
\label{sec:5}
In this work, we introduced a general framework that leverages the coherent superposition of spatially distinct communication links to enable the intrinsic generation of both bipartite and multipartite entanglement during the distribution process. Starting from the bipartite scenario, we demonstrated deterministic generation of Bell states from initially separable inputs, and subsequently extended the approach to the multipartite regime. The quality of the generated states has been assessed through fidelity and concurrence, confirming the effectiveness of the proposed scheme across different configurations.
A key aspect of our analysis is the role of the vacuum amplitudes, which govern the interference effects underlying the superposition of the communication links. We showed that, by properly tuning these parameters, maximally entangled states can be obtained deterministically even in extreme regimes, including entanglement-breaking channels and channels with zero quantum capacity. Our finding challenges the conventional view of noise as purely detrimental, revealing instead that quantum noise, when coherently controlled, can be exploited as a constructive resource for entanglement generation.
Overall, our results establish the coherent superposition of noisy channels as a powerful and experimentally feasible mechanism for entanglement generation, opening new perspectives for quantum communication and distributed quantum network functionalities in realistic noisy environments.


\begin{appendices}
\section{Proof of Propositions ~\ref{prop:01} and ~\ref{prop:02}}
\label{Appendix_A}
We consider the input state $\rho = \rho_t \otimes \rho_c$, where $\rho_t = \ket{0}^{\otimes n}\bra{0}^{\otimes n}$ and $\rho_c = \ket{+_c}\bra{+_c}$. In the noiseless regime, it is convenient to work in the state-vector representation. By applying the operator $S=X^{\otimes n}\otimes\ketbra{0_c}{0_c}+Z^{\otimes n}\otimes\ketbra{1_c}{1_c}$ ruling the superposition, after some algebraic manipulations, Eq.~\eqref{eq:01} becomes:
\begin{align}
\label{eq:20}
&S\left(\ket{0}^{\otimes n}\otimes\ket{+_c}\right)=\\\nonumber
&=\frac{1}{\sqrt{2}}\left(X^{\otimes n}\ket{0}^{\otimes n}\otimes\ket{0_c}+Z^{\otimes n}\ket{0}^{\otimes n}\otimes\ket{1_c}\right)\\\nonumber
&=\frac{1}{2}(\ket{0}^{\otimes n}+\ket{1}^{\otimes n})\ket{+_c}+\frac{1}{2}(\ket{0}^{\otimes n}-\ket{1}^{\otimes n})\ket{-_c})
\end{align}
By measuring the control qubit in the state $\ket{+_c}$($\ket{-_c}$), the target system is projected onto the GHZ state $\ket{GHZ^+}=\frac{1}{\sqrt{2}}(\ket{0}^{\otimes n}+\ket{1}^{\otimes n})$ ($\ket{GHZ^-}=\frac{1}{\sqrt{2}}(\ket{0}^{\otimes n}-\ket{1}^{\otimes n})$. For $n=2$, this reduces to the Bell states $\ket{\Phi^\pm}$. As the two outcomes are locally unitary equivalent, GHZ states (Bell states for $n=2$) are generated deterministically, independently of the measurement outcome.
\section{Proof of Proposition~\ref{prop:03}}
\label{Appendix_B}
We consider the input state $\rho=\rho_t \otimes \rho_c=(\ket{0}\bra{0})^{\otimes n}\otimes\ket{\tilde{0}}\bra{\tilde{0}}$, where $\ket{\tilde{0}}=\frac{1}{\sqrt{n}}\sum_{j=0}^{n-1}\ket{j}$ denotes the uniform superposition in the Fourier basis. By applying the operator $S=\sum^{n-1}_{j=0}X_j\otimes\ket{j}\bra{j}$ to the input state, where $X_j$ are the operators defined in \cref{sec:3.2}, \cref{eq:01} becomes:
\begin{align}
\label{eq:21}
&S\rho S^\dagger=\sum^{n-1}_{l,m=0}(X_l\otimes\ket{l}\bra{l})\,\rho\, (X_m\otimes\ket{m}\bra{m})=\nonumber\\
&=\sum^{n-1}_{l,m=0}(X_l\otimes\ket{l}\bra{l})\left(\ket{0}\bra{0})^{\otimes n}\otimes\ket{\tilde{0}}\bra{\tilde{0}}\right)(X_m\otimes\ket{m}\bra{m})\nonumber\\
&=\frac{1}{n}\sum^{n-1}_{l,m=0}(X_l(\ket{0}\bra{0})^{\otimes n}X_m)\otimes\ket{l}\bra{m}.
\end{align}
By expressing the control qudit in the Fourier basis $\{\ket{\tilde{k}}=1/\sqrt{n}\sum^{n-1}_{l=0}\omega^{k l}\ket{l}\}$ with $\omega=exp(2\pi i/n))$, \eqref{eq:21} becomes:
\begin{equation}
\label{eq:22}
S\rho S^\dagger=\frac{1}{n^2}\sum^{n-1}_{l,m,k,j=0}\omega^{-kl+jm}(X_l(\ket{0}\bra{0})^{\otimes n}X_m)\otimes\ket{\tilde{k}}\bra{\tilde{j}}.
\end{equation}
By measuring the control qudit in the $\ket{\tilde{0}}$ state, the target system is projected onto the state $\rho_W=\ket{W}\bra{W}$ with
\begin{equation}\label{eq:23}
    \ket{W}=\frac{1}{\sqrt{n}}\sum^{n-1}_{l=0}\ket{0_0,0_1,...,1_l,...0_{n-1}}.
\end{equation} 
If the control qudit is measured in any other Fourier basis state, the resulting state is locally equivalent to $\ket{W}$. Thus, the generation of the W state is deterministic.
\section{Fidelity Analysis for Noisy Bell-State Generation}
\label{Appendix_C}   
Consider the output state obtained by coherently superposing two quantum channels $\mathcal{F}(\cdot)$ and $\mathcal{N}(\cdot)$ and subsequently measuring the control qubit in the state $\ket{+_c}$ or $\ket{-_c}$. The corresponding post-measurement state is given by:
\begin{equation}
\label{eq:24}
   \rho'_{\pm} =\sum_{ij} (\beta_j F_i \pm \alpha_i N_j )\rho_t (\beta^*_j F^\dagger_i \pm \alpha^*_i N^\dagger_j).
\end{equation}
For general Pauli channels with noise parameters $\{p_0, p_1, p_2, p_3\}$ and $\{q_0, q_1, q_2, q_3\}$, the non-normalized output state derived from \cref{eq:24} is given in \eqref{eq:25}, reported at the top of this page.
\begin{figure*}
    \begin{align}
    \rho'& =\frac{1}{2}\Big([p_0 + p_3 +q_0+q_3 +\left(\sqrt{p_0} \alpha_0^*+\sqrt{p_3} \alpha_3^*\right) \left({\beta_0} \sqrt{q_0}+{\beta_3} \sqrt{q_3}\right)+({\alpha_0} \sqrt{p_0}+{\alpha_3} \sqrt{p_3})\left(\sqrt{q_0} \beta_0^*+\sqrt{q_3}\beta_3^*\right.)]\ket{00}\bra{00}\nonumber\\
   & +[\left(\sqrt{p_0} \alpha_0^*+\sqrt{p_3}\alpha_3^*\right) \left({\beta_1} \sqrt{q_1}-{\beta_2} \sqrt{q_2}\right)+\left({\alpha_1} \sqrt{p_1}-{\alpha_2} \sqrt{p_2}\right) \left(\sqrt{q_0} \beta_0^*+\sqrt{q_3}\beta_3^*\right) ]\ket{00}\bra{11}\nonumber\\
  & +[ \left({\alpha_0} \sqrt{p_0}+{\alpha_3} \sqrt{p_3}\right) \left(\sqrt{q_1} \beta_1^*-\sqrt{q_2}\beta_2^*\right)+\left(\sqrt{p_1}\alpha_1^*-\sqrt{p_2} \alpha_2^*\right) \left({\beta_0} \sqrt{q_0}+{\beta_3} \sqrt{q_3}\right)]\ket{11}\bra{00}\nonumber\\
   & +[ p_1+p_2+q_1+q_2+\left(\sqrt{p_1} \alpha_1^*-\sqrt{p_2} \alpha_2^*\right) \left({\beta_1} \sqrt{q_1}-{\beta_2} \sqrt{q_2}\right)+\left({\alpha_1} \sqrt{p_1}-{\alpha_2} \sqrt{p_2}\right) \left(\sqrt{q_1}\beta_1^*-\sqrt{q_2} \beta_2^*\right)]\ket{11}\bra{11}\Big).
    \label{eq:25}
\end{align}
\end{figure*}
If the control qubit is measured in the state $\ket{-}\bra{-}$, the corresponding output state is obtained from \cref{eq:25} by inverting the sign of the vacuum amplitudes $\{\alpha_i\}$.

The case of two depolarizing channels is obtained from \cref{eq:25}, by setting $p_0=1-p, p_1=p_2=p_3=p/3$ and $q_0=1-q, q_1=q_2=q_3=q/3$. 
By substituting \cref{eq:25} into the definition of the fidelity $\mathrm{fid}$, namely, $\mathrm{fid}\eqdef \mathrm{Tr}\!\left[\sqrt{\sqrt{\rho'}\,\rho_{\ket{\Phi^+}}\,\sqrt{\rho'}}\right]$, where $\rho_{\ket{\Phi^+}}$ denotes the density matrix of the Bell state $\ket{\Phi^+}$, \cref{eq:07} is obtained. As aforementioned, \cref{eq:07} refers to a measurement of the control qubit in the state $\ket{+_c}\bra{+_c}$, while the expression corresponding to the outcome $\ket{-_c}\bra{-_c}$ is obtained by flipping the signs of $\{\alpha_i\}$. 

A relevant specialization is obtained by restricting to the case with noise parameters $p_2=p_3=0$, $q_1=q_2=0$, and $p_0=1-p$, $p_1=p$, $q_0=1-q$, $q_3=q$, corresponding to the two-qubit correlated extensions of phase-flip and a bit-flip channels. The resulting output state obtained through the spatial superposition takes the form:
\begin{align}
\label{eq:26}
   \rho' = K\Big(&\big[\,2-p
+(\alpha_0\beta_0^*+\alpha_0^*\beta_0)\sqrt{(1-p)(1-q)}
\nonumber\\&+ (\alpha_0\beta_3^*+
\alpha_0^*\beta_3)\sqrt{(1-p)q}\,\big]\ket{00}\bra{00}
\nonumber\\
&+\big[\,\alpha_1\beta_0^*\sqrt{p(1-q)}+\alpha_1\beta_3^*\sqrt{pq}\,\big]\ket{00}\bra{11}
\nonumber\\
&+\big[\,\alpha_1^*\beta_0\sqrt{p(1-q)}+\alpha_1^*\beta_3\sqrt{pq}\,\big]\ket{11}\bra{00}
\nonumber\\
&+p\,\ket{11}\bra{11}\Big),
\end{align}
with normalization factor $
K=1/[2+2\Re(\alpha_0\beta_0^*)\sqrt{(1-p)(1-q)}
+2\Re(\alpha_0\beta_3^*)\sqrt{(1-p)q}]$. As in the depolarizing case, by substituting \eqref{eq:26} into the definition of the fidelity, one obtains Eq.~\eqref{eq:27}, reported at the top of the next page.
\begin{figure*}[ht]
\begin{equation}
\label{eq:27}
    \mathrm{fid}=\sqrt{\frac{1+\sqrt{q(1-p)}\alpha_0\beta_3+\sqrt{(1-p)(1-q)}\alpha_0\beta_0+\sqrt{p(1-q)}\alpha_1\beta_0+\sqrt{pq}\alpha_1\beta_3}{2(1+\sqrt{q(1-p)}\alpha_0\beta_3+\sqrt{(1-p)(1-q)}\alpha_0\beta_0)}}.
\end{equation}
\end{figure*}
In the expressions for the fidelity, we have assumed that the vacuum coefficients are real. In the case of complex vacuum coefficients, the term $\alpha_i \beta_j$ should simply be replaced by $(\alpha^*_i\beta_j+\alpha_i\beta_j^*)/2$.

\section{Proof of Proposition~\ref{prop:04} and Corollary~\ref{cor:01}}
\label{Appendix:D}
Consider the fidelity in \cref{eq:07} with coefficients $C$ and $D$ given by Eqs.~\eqref{eq:08} and \eqref{eq:09}, for $p=q$. By evaluating the resulting expression for the two relevant noise regimes, namely $p=q=1$ and $p=q=1/2$, and imposing the condition $\mathrm{fid}=1$, one obtains a set of algebraic constraints on the vacuum amplitudes. Solving these constraints under the normalization conditions $\sum_i |\alpha_i|^2=1$ and $\sum_j |\beta_j|^2=1$ yields the vacuum configurations reported in \cref{eq:10} and \cref{eq:11}, thus proving Proposition~\ref{prop:04}.

A similar procedure applies to the bit-flip/phase-flip case. By evaluating the fidelity in \cref{eq:27} for $p=q=1$ and $p=q=1/2$, the condition $\mathrm{fid}=1$ leads to the corresponding constraints on the vacuum amplitudes. The solutions of these constraints, reported in \cref{eq:12} and \cref{eq:13}, prove Corollary~\ref{cor:01}.

One can observe that the vacuum amplitudes are optimized with respect to the $\ket{+_c}\bra{+_c}$-measurement outcome, as the fidelity expressions used to impose the unit-fidelity condition are derived for a projection onto $\ket{+_c}\bra{+_c}$. However, with this choice of vacuum coefficients, measuring the control qubit in the state $\ket{-_c}\bra{-_c}$ only introduces a relative phase (see Appendix~\ref{Appendix_C}), corresponding to a local unitary transformation. Hence, both outcomes yield maximally entangled states, ensuring deterministic entanglement generation up to local corrections.

\section{Proof of Proposition~\ref{prop:05} and Corollary~\ref{cor:02}}
\label{Appendix_E}
Proceeding analogously to the Bell-state case, one finds that the fidelity expression in Eq.~\eqref{eq:07} also applies to the GHZ-state scenario. In particular, Eq.~\eqref{eq:07} holds exactly when $n \equiv 0 \ (\mathrm{mod}\ 4)$. Importantly, this restriction does not represent a limitation of the framework, but rather a key feature. Specifically, instead of tailoring the optimization of the vacuum amplitudes to the parity of $n$, we adopt a parity-independent configuration. This is possible because for $n \not\equiv 0 \ (\mathrm{mod}\ 4)$, the generated state, obtained using the vacuum coefficients optimized for the case $n \equiv 0 \ (\mathrm{mod}\ 4)$, differs from the standard GHZ state only by a relative phase originating from the factor $i^n$ associated with the $Y^{\otimes n}$ contribution. This phase can always be compensated by a local unitary operation at the receiver side. 

Therefore, optimizing the vacuum coefficients under the assumption $n \equiv 0 \ (\mathrm{mod}\ 4)$ allows one to avoid reconfiguring the setup depending on the system size. This in turn yields a more robust and scalable implementation, while preserving deterministic generation of maximally entangled states up to local corrections.

By specializing the fidelity expression to the cases  $p=q=1$ and $p=q=1/2$, one has the vacuum configurations reported in Proposition~\ref{prop:05}.

Similarly, applying the same procedure to the fidelity expression for the bit- and phase-flip case yields the vacuum amplitudes reported in Eqs.~\eqref{eq:17} and \eqref{eq:18}, thus proving Corollary~\ref{cor:02}.
\section{Proof of Proposition~\ref{prop:07}}\label{Appendix_F}
It's trivial to see that the output state of the superposition of the proposed channels is
\begin{equation}\label{eq:31}
\rho'=\frac{1}{n}\sum^1_{i_0,...,i_{n-1}=0}\sum^{n-1}_{l,m}\prod_{k\neq l}\prod_{j\neq m}\alpha^{(k)}_{i_k}\alpha^{*(j)}_{i_j}E^{(m)}_{i_m}\rho_t E^{\dagger(l)}_{i_l}. 
\end{equation}
Evaluating the fidelity for the case $n=3$ (the $n$ qubit case can be straightforwardly extended), we find the expression given in \eqref{eq:35}.
\begin{figure*}[!t]
\begin{align}
\label{eq:35}
\mathrm{fid}=\sqrt{\frac{p_1+p_2+p_3+2\alpha^{(1)}_1\alpha^{(2)}_1\sqrt{p_1 p_2}+2\alpha^{(2)}_1\alpha^{(3)}_1\sqrt{p_2 p_3}+2\alpha^{(1)}_1\alpha^{(3)}_1\sqrt{p_1 p_3}}{9+6\alpha^{(1)}_0\alpha^{(2)}_0\sqrt{(1-p_1)(1- p_2)}+6\alpha^{(2)}_0\alpha^{(3)}_0\sqrt{(1-p_2)(1- p_3)}+6\alpha^{(1)}_0\alpha^{(3)}_0\sqrt{(1-p_1)(1- p_3)}}}.
\end{align}
\end{figure*}
It is easy to see from \cref{eq:35} that again if $\mathrm{fid}$ is put equal to $1$ solving it for the vacuum amplitudes  values one find that the solutions are $\alpha^{(2)}_{1}=\alpha^{(0)}_1 =\alpha^{(1)}_1=1$.
\end{appendices}

\bibliographystyle{IEEEtran}
\bibliography{Bibliography.bib}

\end{document}